\documentclass[12pt]{article}

\usepackage[font=small,format=plain,labelfont=bf,textfont=it]{caption}
\usepackage{setspace}
\usepackage{apptools}

\usepackage{fullpage, verbatim, amsfonts, amsmath, amssymb, amsthm, latexsym, setspace, graphicx}
\usepackage[T1]{fontenc}
\usepackage[latin9]{inputenc}
\usepackage{geometry}
\usepackage{xcolor}
\usepackage{natbib}
\usepackage{overpic}

\usepackage[french, english]{babel}
\usepackage{hyperref,bookmark}

\hypersetup{linkcolor=MyDarkBlue, citecolor=MyDarkBlue, colorlinks=true} 
	\definecolor{MyDarkBlue}{rgb}{0,0.08,0.45}
	\definecolor{MyDarkGreen}{rgb}{0,0.55,0.08}

\theoremstyle{plain}
\newtheorem{thm}{Theorem}
\newtheorem{lem}{Lemma}
\newtheorem{cor}{Corollary}
\newtheorem{prop}{Proposition}

\newtheorem{assume}{Assumption}

\newtheorem{ex}{Example}

\theoremstyle{definition}
\newtheorem{rem}{Remark}

\AtAppendix{\counterwithin{lem}{section}}
\AtAppendix{\counterwithin{prop}{section}}
\AtAppendix{\counterwithin{thm}{section}}
\AtAppendix{\counterwithin{ex}{section}}

\newcommand{\argmin}{\mathop{\mathrm{argmin}}}
\newcommand{\bE}{{\mathbb E}}

\newcommand\cites[1]{\citeauthor{#1}'s \citeyearpar{#1}}

\newcommand{\finex}{\leavevmode\unskip\penalty9999 \hbox{}\nobreak\hfill\quad\hbox{$\blacktriangle$}}

\begin{document}
\title{Monitoring with Rich Data\thanks{Frick: Princeton University (mira.frick@gmail.com); Iijima: Princeton University (iijimaaa@gmail.com); Ishii: Pennsylvania State University (yxi5014@psu.edu). For helpful comments, we thank Nageeb Ali, Dirk Bergemann, Alex Bloedel, Hector Chade, Drew Fudenberg, George Georgiadis, Nima Haghpanah, Marina Halac, Johannes H\"{o}rner, Michihiro Kandori, Vijay Krishna, Giacomo Lanzani, Yingkai Li, Akihiko Matsui, Larry Samuelson, Ran Shorrer, Philipp Strack, Takuo Sugaya, Bal\'azs Szentes, Alex Wolitzky, and numerous seminar audiences. We thank Haoning Chen for excellent research assistance. Frick gratefully acknowledges the financial support from a Sloan Research Fellowship. }}

	\author{Mira Frick \and Ryota Iijima \and Yuhta Ishii}
\date{First posted version: April 8, 2023 \\ \vspace{3mm} This version: \today}
\maketitle 

\begin{abstract}

We offer a new perspective on the issue that commonly used contracts tend to be simple, even though standard models predict more complicated optimal contracts. We consider moral hazard problems where a principal has access to rich monitoring data about an agent's action. Rather than focusing on optimal contracts, we measure the performance of contracts by analyzing the rate at which the principal's payoffs converge to the first-best as the amount of data grows large. Our main result shows that the optimal convergence rate to the first-best is achieved by binary wage schemes, suggesting a novel rationale for this simple and widely observed class of contracts. Notably, in order to attain the optimal convergence rate, the principal must set a lenient cutoff for when the agent receives a high vs.\ low wage. In contrast, we find that other common contracts where wages vary more finely with observed data (e.g., linear contracts) approximate the first-best at a highly suboptimal rate. Finally, we show that the optimal convergence rate depends only on a simple summary statistic of the monitoring technology. This yields a detail-free ranking over monitoring technologies that quantifies their value for incentive provision in data-rich settings and applies regardless of the agent's specific utility or cost functions.

\medskip


\end{abstract}

\onehalfspacing

\newpage

\section{Introduction}
\subsection{Motivation and Overview}

An enduring issue in contract theory is that wage schemes observed in practice are often ``simple,'' even though textbook models yield more complicated optimal contracts.\footnote{See, e.g., the recent survey by \cite{georgiadis2022}.} In this paper, we suggest a novel perspective on this problem. We consider a standard static moral hazard setting, where a principal (e.g., employer) designs a wage scheme to incentivize an agent (e.g., worker) whose costly action choice is not directly observable to the principal. As a key feature, we assume that the principal has access to fairly \emph{rich/precise data} about the agent's action, as may be the case in many workplaces (e.g., due to modern technologies such as automated quality control systems or productivity tracking software, but also more traditional monitoring tools such as customer or co-worker evaluations). If the principal is able to perfectly monitor the agent's action, there is no moral hazard and the principal can achieve her first-best payoff.  Away from this extreme limit, we are interested in which contracts allow the principal to efficiently exploit rich but imperfect monitoring data, and whether, and if so which, simple contracts may be enough for this purpose.

The main novelty of our approach is that, to capture the efficient exploitation of rich data, we do not focus on which contracts are optimal (i.e., maximize the principal's payoff). Rather, we analyze the \emph{rate of convergence} of the principal's payoff to the first-best as the amount of data grows rich, and we focus on the less demanding criterion of which contracts achieve the \emph{optimal convergence rate}. The convergence rate to the first-best is a useful measure of a contract's performance in data-rich settings: Contracts with a higher convergence rate yield higher payoffs for the principal than contracts with a lower convergence rate whenever data is rich enough. Moreover, while optimal contracts necessarily converge to the first-best at the optimal rate, there may be simpler and more realistic classes of contracts that achieve the same optimal convergence rate.

Indeed, our main result is that the optimal convergence rate is achieved by a particular class of simple contracts that are widely used in practice: \emph{binary} payment schemes, i.e., contracts with only two possible wage levels. Notably, in order to attain the optimal convergence rate, the principal must set a \emph{maximally lenient} cutoff for when the agent receives the high vs.\ low wage. In contrast, we find that other common contracts where wages vary more finely with observed data (e.g., linear contracts) approximate the first-best at a highly suboptimal rate. Finally, we show that the optimal convergence rate depends only on a simple summary statistic of the monitoring technology. This yields a detail-free ranking over monitoring technologies that quantifies their value for incentive provision in data-rich settings and applies regardless of the agent's specific utility or cost functions.

In our baseline model (Section~\ref{sec:model}), the principal has access to a monitoring technology that generates $n$ signals about the agent's action $a$, where $n$ parametrizes the richness/precision of the principal's data. For example, $a$ may represent a factory worker's assembly of a widget, and upon completion $n$ signals are generated by an automated quality control system that repeatedly scans the widget for various errors; or $a$ may represent an instructor's effort, and signals take the form of reviews submitted by the $n$ students at the end of the course. Our baseline model assumes that the agent's choice of $a$ is one-shot and signals are generated i.i.d.\ conditional on $a$, but Section~\ref{sec:general} extends the analysis to more general settings. The principal seeks to implement a target action by designing a contract that specifies a wage payment contingent on each realized signal sequence, subject to standard individual rationality (IR) and incentive compatibility (IC) constraints. The principal is risk-neutral while the agent is risk-averse. We study the optimal rate at which the principal's implementation cost converges to the first-best (i.e., observable action case) as the amount of data $n$ grows large.

Our main result (Theorem~\ref{thm:main}) shows that, regardless of whether the principal optimizes over general contracts or binary contracts, her payoffs converge to the first-best at the same exponential rate. This optimal convergence rate is given by the Kullback-Leibler (KL) divergence between the signal distribution under the target action and the closest signal distribution under any less costly deviation. Thus, this rate depends only on the detectability of the hardest-to-detect deviation from the target action.

The binary contracts that achieve the optimal convergence rate take the form of statistical tests: The principal uses a score to partition signal sequences into a ``pass'' (high wage) and ``fail'' (low wage) region. Importantly, the convergence rate depends on the choice of the pass--fail cutoff.  Choosing the cutoff entails a trade-off between the risk of {\it false negatives} (i.e., failing the test under the target action) and {\it false positives} (i.e., passing the test under a deviation). 
We show that false negatives become the dominant source of inefficiency as the amount of data grows large. Thus, in order to achieve the optimal convergence rate, the principal must make the cutoff as lenient as possible subject to IC.

Binary wage schemes are frequently used in practice.  A prominent example are single-bonus contracts (a base wage, plus a fixed bonus if performance is sufficiently good), which are common in many professions.\footnote{Another example highlighted by the literature are jobs that feature essentially flat wages but entail the threat of being fired in case of sufficiently poor performance. See, e.g., \cite{murphy1999}, \cite{georgiadis2020}, and references therein.} 
Complementing existing work that has focused on identifying settings in which binary wage schemes are exactly optimal (see Section~\ref{sec:lit}), Theorem~\ref{thm:main} suggests a novel perspective on the use of such contracts: Provided a principal has access to rich enough data about workers' actions, any benefit from optimizing over more complex, non-binary wage schemes becomes negligible, as both general contracts and binary contracts approximate the first-best at the same rate. In this sense, even though binary contracts are suboptimal, these simple wage schemes are an effective way for the principal to exploit rich monitoring data. Our finding of leniency may also conform with evidence about binary contracts used in practice: For example, in the context of single-bonus contracts, \cite{joseph1998} highlight that organizations tend to use bonuses to reward ``acceptable'' rather than ``exceptional'' performance.\footnote{They conduct a survey of how selective firms that use bonus contracts are about whether or not to pay workers the bonus. The modal response is the least selective option (i.e., award a bonus to 75-100\% of workers). See also \cite{johnston2016}.}

To illustrate that contracts with more fine-grained wage variation can approximate the first-best at a highly suboptimal rate, Proposition~\ref{prop:linear} considers another widely observed class of contracts---linear wage schemes. Under these contracts, we show that the convergence to the first-best is much slower than optimal, viz.\ subexponential. Thus, compared with binary contracts, linear contracts perform quite poorly at exploiting rich data, and for any large enough $n$, the principal is better off using binary than linear contracts. Proposition~\ref{prop:limit2} further adds to the rationale for binary contracts in data-rich settings, by showing that, as $n$ grows large, any sequence of optimal contracts approximates a binary contract of the maximally lenient form we identify above.

Our characterization of the optimal convergence rate in Theorem~\ref{thm:main} also yields a ranking over monitoring technologies that quantifies their value for incentive provision (Corollary~\ref{cor:ranking}): In data-rich settings, monitoring technologies with higher KL divergence between the signal distributions under target vs.\ non-target actions are more valuable to the principal than those with lower KL divergence. Notably, this ranking is independent of the 
agent's utility over money and specific cost function, providing detail-free guidance to a principal choosing between different monitoring technologies.

Section~\ref{sec:discussion} discusses some variants of our baseline model. First, Section~\ref{sec:limliab} introduces a severe form of limited liability under which IR cannot bind. We show that binary contracts continue to achieve the optimal convergence rate to the first-best, but the rate of convergence is slower and based on a different statistical measure, reflecting a tradeoff between risk aversion and rent extraction that is absent in the main model. Second, beyond our baseline model with $n$ i.i.d.\ signal draws about a one-shot action, Section~\ref{sec:general} extends our main result to two formulations of rich monitoring data that allow for (i) general sequences of increasingly precise monitoring technologies (e.g., vanishing observation noise, serially correlated data) or (ii) repeated but infrequent action adjustments.

Finally, at a methodological level, our analysis highlights that, in settings where optimal contracts are complicated, convergence rates to the first-best can be a useful measure to quantify and compare the performance of simple but suboptimal contracts. Focusing on convergence rates reduces the evaluation of contracts to a one-dimensional statistic that does not depend on all the specifics of the environment (e.g., the agent's preferences or the fine details of the monitoring technology); this permits sharp predictions and allows us to draw on tools from large deviation theory. At the same time, by its nature, the convergence rate approach is silent about outcomes at fixed $n$, which would depend on the specifics of the environment. This  may raise the concern that our results might only apply at unrealistically large $n$, at which point any performance gap between the classes of contracts we study may be economically uninteresting, as all may yield payoffs very close to the first-best. To mitigate this concern, it is helpful to complement the convergence rate results with an analysis of specific environments where the performance of different contracts can be studied analytically as a function of $n$. In Example~\ref{ex:finite-n}, we illustrate this for a setting with binary actions, Gaussian signals, and CARA utilities. Under natural parameters, lenient binary contracts start to come close to the first-best at quite moderate $n$ and outperform linear contracts by a significant margin, reinforcing the predictions of the convergence rate analysis.

\subsection{Related Literature}\label{sec:lit}

In the standard moral hazard setting \`a la \cite{holmstrom1979} with a risk-averse agent, binary contracts are optimal only under restrictive, binary monitoring technologies (i.e., whose distribution of signal likelihood ratios has binary support).\footnote{With a risk-neutral agent, binary contracts can be optimal under various other assumptions \citep[e.g.,][]{oyer2000, palomino2003, levin2003}, but these findings rely on exact risk neutrality.} Several papers derive the optimality of binary contracts by enriching this setting. For example, \cite{georgiadis2020} consider a principal who can flexibly design a monitoring technology at a cost.\footnote{Specifically, the principal chooses when to stop observing the output of a diffusion process whose drift is the agent's action, at a cost proportional to the principal's stopping time.} They show that the principal optimally chooses a binary monitoring technology, by establishing a connection with an information design problem. \cite{herweg2010} and \cite{lopomo2011} instead consider agents with non-expected utility preferences. Complementary to these papers, we revisit a standard moral hazard setting with an exogenous monitoring technology and risk-averse expected-utility agent, but we relax the criterion of exact optimality. Instead, we provide a rationale for binary contracts based on the idea that in data-rich settings, they allow the principal to approximate the first-best at the optimal rate. We also highlight the importance of a lenient high--low wage cutoff for achieving the optimal convergence rate, which is not a general feature of the optimal binary contracts in the aforementioned settings.\footnote{\cite{herweg2010} highlight that the set of signals that result in a high wage sometimes (but not always) contains some ``bad'' signals that are more indicative of lower effort.} Several papers identify natural forces that favor linear contracts \citep[e.g.,][]{holmstrom1987, carroll2015, barron2020, walton2022}. In contrast, we show that in our data-rich setting, linear contracts perform significantly worse than binary contracts.

Convergence rates as a measure of the performance of simple but suboptimal mechanisms have also been analyzed in other settings. For example, in the context of large markets, a classic literature studies the rate at which simple trading mechanisms approximate efficiency as the number of market participants grows large.\footnote{See, e.g., \cite{rustichini1994, satterthwaite2002, hong2004, pakzad2023}. In other settings, some papers show that simple mechanisms can approximate the first-best, without analyzing convergence rates \cite[e.g.,][in the context of dynamic moral hazard problems where the discount factor grows large]{radner1985}.} While much of this literature does not focus on identifying mechanisms that achieve the optimal convergence rate, \cite{satterthwaite2002} show that double auctions achieve the optimal worst-case convergence rate (i.e., when evaluated at the least favorable trading environment). In contrast, we find that binary contracts can be used to achieve the optimal convergence rate in all moral hazard environments we consider, not just the worst-case environment.\footnote{Relatedly, our more recent paper \citep{FII-screening} studies multi-good monopoly and shows that a simple selling mechanism---pure bundling---allows the seller to approximate the first-best revenue at the optimal rate as her data about buyers' valuations grows rich.} The optimal convergence rate to the first-best is not directly comparable to a different notion of approximate optimality that is often analyzed in the computer science literature: worst-case guarantees, i.e., lower bounds on the performance ratio of simple vs.\ optimal mechanisms that are uniform with respect to some features of the environment \citep[for a survey, see][]{roughgarden2019}. In moral hazard environments, \cite{dutting2019} obtain such a guarantee for linear contracts.

Our ranking over monitoring technologies (Corollary~\ref{cor:ranking}) relates to the literature on the value of monitoring in moral hazard problems \citep[e.g.,][]{holmstrom1979, gjesdal1982, singh1985, kim1995, jewitt2007}. Several papers derive partial orders over monitoring technologies in settings where the principal observes a single signal draw about the agent's action. As we discuss in Section~\ref{sec:ranking}, by quantifying the convergence rate to the first-best as the amount of data grows large, we obtain a ranking over monitoring technologies that is a completion of the orders in \cite{kim1995} and \cite{jewitt2007}. As such, Corollary~\ref{cor:ranking} is an analog for incentive problems of \cite{moscarini2002}: They quantify the value of rich data in learning settings, based on a different index that characterizes the rate at which a decision-maker who observes repeated i.i.d.\ signal draws from an information structure learns the state; this yields a completion of \cites{blackwell} order over information structures.\footnote{Convergence rates are also analyzed to measure efficiency in other learning settings, including social learning \citep[e.g.,][]{vives1993, harel2021}, higher-order beliefs \citep[e.g.,][]{frick2022}, and misspecified learning \citep[e.g.,][]{FII21}.} Comparisons of monitoring structures have also been studied in the context of repeated games \citep[e.g,][]{kandori1992}. Some recent papers analyze how the monitoring structure affects the convergence rate of equilibrium payoffs to the efficient frontier as players become patient \citep{horner2016, sugaya2022, sugaya2023}.

\section{Model}\label{sec:model}

{\noindent {\bf Environment.}} Consider the following static moral hazard setting. There is a principal (``she'') and an agent (``he''). The agent chooses a one-shot action from a finite action set $A$.\footnote{We do not impose any order assumptions on $A$; for example, actions may be multi-dimensional.} The principal does not observe the agent's action, but has access to a monitoring technology $\mu$: This specifies a signal space $X$, assumed to be a subset of a Euclidean space, and, for each chosen action $a$, a distribution $\mu_a \in \Delta(X)$ over signals, where $\Delta(X)$ denotes the set of Borel probability measures over $X$. 

For all distinct actions $a \neq a'$, we impose the identification assumption that the signal distributions $\mu_a$ and $\mu_{a'}$ are different. We also assume that $\mu_{a'}$ is absolutely continuous with respect to $\mu_{a}$ (with corresponding Radon-Nikodym derivative $\frac{d\mu_{a'}}{d\mu_{a}}$), which implies that no signals perfectly reveal the chosen action. Finally, we impose the regularity condition that the moment-generating functions of signal log-likelihood ratios are well-defined, i.e., $\int \left(\frac{d\mu_{a'}}{d\mu_a}(x)\right)^\lambda d\mu_a(x)<\infty$ for all $\lambda> 0$.

To capture that the principal has access to rich data about the agent's action, we assume that after the agent chooses his action $a$, the principal observes $n$ i.i.d.\ draws of signals $x^n=(x_1,\ldots, x_n)$ from $\mu_a$. Here, $n$ parametrizes the richness or precision of the principal's data, and we will be interested in settings where $n$ is large. Let $\mathbb{P}_a$ denote the distribution over signal sequences $x^n$ conditional on action $a$; let ${\mathbb E}_a$ and ${\rm Var}_a$ denote the corresponding expectation and variance operators. 

Our assumption that the agent chooses a one-shot action which then generates multiple i.i.d.\ signals can be seen as an approximation of some real-world settings with rich monitoring data, such as the automated quality control or teaching evaluation examples mentioned in the Introduction. However, as Section~\ref{sec:general} discusses, the analysis generalizes readily to less stylized settings: First, moving beyond the i.i.d.\ signal formulation, we can consider general sequences $\mu^n$ of increasingly precise monitoring technologies (e.g., one-shot observation of the action perturbed by some vanishing noise, or repeated but serially correlated signals). Second, moving beyond one-shot action choice, we can allow the agent to repeatedly adjust his action, as long as action adjustments are infrequent relative to the number $n$ of signals.

\medskip

{\noindent {\bf Payoffs.}} The principal is risk-neutral and seeks to implement some target action $a^* \in A$ by designing a contract, i.e., a wage scheme $w: X^n \to \mathbb  [\underline w, \infty)$ that specifies a payment $w(x^n)$ contingent on each realized signal sequence $x^n$.\footnote{Our results extend readily to the case of a risk-averse principal. As is common in the literature, we focus on the principal's incentive design problem and do not explicitly model her optimization over the target action $a^*$, but incorporating the latter is straightforward (for instance, if there is a unique first-best optimal target action $a^*$, then the optimal target action under all classes of contracts we consider coincides with $a^*$ at large enough $n$).} To rule out ``shoot the agent'' arguments \`a la \cite{mirrlees1999} (see Remark~\ref{rem:mirrlees}), we impose a lower bound $\underline w\in\mathbb R$ on wages, which can be arbitrarily low.

 The agent's payoffs are additively separable, consisting of a consumption utility $u : [\underline w, \infty)\to\mathbb R$ over money minus a cost $c: A\to\mathbb R$ for each action. The agent has an outside option, whose payoff we normalize to $0$. We assume the following: 
 \begin{assume}\label{asp} \
\begin{enumerate} 
\item $u$ is twice continuously differentiable with $u' (w) > 0$ and $u'' (w) < 0$ for all $w \in [\underline w, \infty)$ and $\lim_{w \to \infty} u'(w) = 0$;

\item $c(A)\subseteq {\rm int} (u([\underline w, \infty)))$;

\item  $c(a^*) > c(a)$ for some $a \in A$ and $c(a^*) \neq c(a)$ for all $a \in A \setminus \{a^*\}$. 
\end{enumerate}
 \end{assume}

The first condition requires the agent to be strictly risk-averse, which will play an essential role in our analysis.\footnote{\label{fn:risk-neutral}Assumption~\ref{asp}.1 is satisfied under standard families of utility functions, e.g., CARA or CRRA. If instead $u$ is linear, standard arguments imply that the principal can achieve the first-best provided there is an (\ref{eq:IC}) contract that satisfies (\ref{eq:IR}) with equality. Under Assumption~\ref{asp}.2, this is the case provided either the range of $u$ is large enough or $n$ is large enough.} Throughout, we let $h := u^{-1}$. The second condition requires the consumption utility range to be rich enough that (under perfect monitoring) suitable wage payments can make the the agent indifferent between each action and the outside option; Section~\ref{sec:limliab} considers the case when this condition is violated. To avoid trivialities, the third condition assumes that the target action $a^*$ is not the least costly action.

 \medskip
\noindent{{\bf Second-best problem.}} In the optimal contract, the principal chooses a wage scheme $w$ to minimize the implementation cost of $a^*$,\footnote{At all large enough $n$, the $\inf$ in (\ref{eq:SB}) is attained by some wage scheme; see Appendix~\ref{app:optimal}.}
\begin{equation}\label{eq:SB}
C^{\rm{SB}}_n(\mu, u, c, a^*)=\inf_{w:X^n\to [\underline w, \infty)} {\mathbb E}_{a^*}[w(x^n)],
\end{equation}
subject to standard incentive compatibility (IC) and individual rationality (IR) constraints:
\begin{equation}\tag{IC}\label{eq:IC}
\bE_{a^*}[u(w(x^n))]-c(a^*) \geq \bE_{a}[u(w(x^n))]-c(a), \quad \forall a \in A,
\end{equation}
\begin{equation}\tag{IR}\label{eq:IR}
\bE_{a^*}[u(w(x^n))]-c(a^*)\geq 0.
\end{equation}
We refer to the induced minimal cost $C^{\rm{SB}}_n(\mu, u, c, a^*)$ as the \textit{\textbf{second-best cost}}. As is standard, it is convenient to rewrite the principal's problem in utility terms. That is, instead of choosing wage schemes $w$, the principal equivalently chooses maps $v:X^n\to u([\underline w, \infty))$ from signal sequences to consumption utilities,
\begin{equation*}
C^{\rm{SB}}_n(\mu, u, c, a^*)=\inf_{v:X^n\to u([\underline w, \infty))} {\mathbb E}_{a^*}[h(v(x^n))]
\end{equation*}
subject to the IC and IR constraints
\begin{equation}\tag{IC}\label{eq:IC'}
 {\mathbb E}_{a^*}[v(x^n)]-c(a^*) \geq   {\mathbb E}_a[v(x^n)]-c(a), \quad \forall a \in A,
\end{equation}
\begin{equation}\tag{IR}\label{eq:IR'}
 {\mathbb E}_{a^*}[v(x^n)]-c(a^*)\geq 0.
\end{equation}

\medskip
\noindent{{\bf Convergence rate analysis.} As is well-known, the second-best problem gives rise to wage schemes that depend finely on realized signals, in a potentially complicated manner that does not in general resemble contracts observed in practice (e.g., binary or linear contracts). Thus, instead of focusing on the second-best cost for a given $n$, we analyze the convergence rate of the second-best cost to the first-best as $n$ grows large. Formally, let $C^{\text{FB}}(u,c, a^*)$ denote the \textit{\textbf{first-best cost}}, i.e., the minimal cost of implementing $a^*$ under perfect monitoring. Under Assumption~\ref{asp}.2, this is the same as the minimal implementation cost of $a^*$ when only (\ref{eq:IR}) is imposed and is given by 
$$C^{\text{FB}}(u,c, a^*)=h(c(a^*)).$$

Clearly, with infinitely many signals, the second-best cost coincides with the first-best, $C^{\rm{SB}}_\infty(\mu, u,c,a^*)=C^{\text{FB}}(u,c,a^*)$, as $n = \infty$ corresponds to perfect monitoring given the identification assumption on $\mu$. Instead, we are interested in environments where the amount $n$ of data is large but finite. To understand how well the principal can perform in such settings, we will characterize the rate at which $C^{\rm{SB}}_n(\mu,u,c,a^*)$ converges to $C^{\text{FB}}(u,c,a^*)$ as $n\to\infty$, i.e., the optimal rate at which the principal can approximate the first-best.

This will allow us to address two questions: First, are there simple classes of contracts that attain this same optimal convergence rate? As noted, achieving the optimal convergence rate to the first-best is a natural criterion for efficient exploitation of rich data: Whenever $n$ is large enough, contracts with a higher convergence rate yield a lower implementation cost than contracts with a lower convergence rate. Second, which monitoring technologies are more valuable for incentive provision in data-rich settings, i.e., how does the optimal rate of convergence vary across $\mu$?

\section{Analysis}

\subsection{Convergence Rates}

{\noindent {\bf Optimal convergence under binary contracts.}} Our main result characterizes the rate at which the second-best cost converges to the first-best as the amount of data grows large. Moreover, we show that this optimal rate of convergence can be achieved by a particular class of simple contracts: \textit{\textbf{binary contracts}}, i.e., wage schemes $w$ with $|w(X^n)|= 2$. 

Denote by $C^{{\rm bin}}_n(\mu, u,c, a^*)$ the second-best cost when, instead of optimizing over all IC and IR wage schemes, the principal is restricted to optimizing over IC and IR binary contracts. Let $A^{-} (c, a^*) := \{ a \in A: c(a) < c(a^*)\}$ be the set of actions that are less costly to the agent than the target action $a^*$, which is nonempty by Assumption~\ref{asp}.3. For all $\nu,\nu'\in\Delta(X)$, denote by ${\rm KL}(\nu, \nu')$ the Kullback-Leibler (KL) divergence of $\nu$ relative to $\nu'$,  i.e.,
\[
{\rm KL}(\nu, \nu'):=
\begin{cases}
\int \ln\left(\frac{d\nu}{d\nu'}(x)\right) \, d\nu(x) & \text{ if } \nu \text{ is absolutely continuous w.r.t. } \nu' \\
\infty &  \text{ otherwise. }
\end{cases}
\]

\begin{thm}\label{thm:main} Under both general and binary contracts, the second-best cost converges to the first-best exponentially at rate $\min_{a \in A^{-} (c, a^*)}{\rm KL} (\mu_a, \mu_{a^*})$: We have
\begin{equation}\label{eq:general-rate}
C^{\rm{SB}}_n(\mu, u,c, a^*)-C^{\rm{FB}}(u,c, a^*)=\exp[-\min_{a \in A^{-} (c, a^*)}{\rm KL} (\mu_a, \mu_{a^*})n+o(n)];
\end{equation}
\begin{equation}\label{eq:binary-rate}
C^{{\rm bin}}_n(\mu,  u,c, a^*)-C^{\rm{FB}}(u,c, a^*)=\exp[-\min_{a \in A^{-} (c, a^*)}{\rm KL} (\mu_a, \mu_{a^*})n+o(n)].
\end{equation}
\end{thm}

By (\ref{eq:general-rate}), when the principal optimizes over all IC and IR wage schemes, she approximates the first-best at an exponential rate given by $\min_{a \in A^{-} (c, a^*)}{\rm KL} (\mu_a, \mu_{a^*})$.\footnote{By the assumptions on $\mu$, ${\rm KL} (\mu_a, \mu_{a^*}) \in (0, \infty)$ for all $a \neq a^*$, so this rate is positive and finite.} To interpret, note that ${\rm KL}(\mu_a, \mu_{a^*})$ is a statistical measure that quantifies how dissimilar the signal distribution under a deviation to action $a \neq a^*$ is from the signal distribution under the target action $a^*$. Thus, (\ref{eq:general-rate}) shows that, for any given monitoring technology $\mu$, the optimal rate of convergence to the first-best depends only on a simple statistic: the detectability $\min_{a \in A^{-} (c, a^*)}{\rm KL} (\mu_a, \mu_{a^*})$ of the hardest-to-detect deviation to a less costly action.

Crucially, (\ref{eq:binary-rate}) implies that the principal can achieve this same optimal rate of convergence using binary contracts. Thus, Theorem~\ref{thm:main} offers a novel rationale for this simple and common class of contracts: While for any given $n$, binary contracts are in general suboptimal, Theorem~\ref{thm:main} shows that binary contracts are an effective way to exploit rich data. As the amount of data $n$ grows large, any benefit from using more general, non-binary contracts has at most a second-order effect on the convergence to the first-best: The only potential difference between (\ref{eq:general-rate}) and (\ref{eq:binary-rate}) is in the terms $o(n)$, but as these are sublinear (i.e., $\lim_{n\to\infty}\frac{o(n)}{n}=0$) they become negligible at large $n$. In terms of the principal's payoffs, this can be quantified by saying that at large $n$, the benefit of optimizing over general vs.\ binary contracts is smaller than the benefit of having access to even a vanishingly small fraction $\varepsilon$ of additional signals.\footnote{That is, for any $\varepsilon > 0$, Theorem~\ref{thm:main} implies that $C^{{\rm bin}} _{\lceil(1+\varepsilon)n\rceil} (\mu,  u,c, a^*)<C^{\rm{SB}}_{n  }(\mu,  u,c, a^*)$ for all large enough $n$. This is similar in flavor to \cites{bulow1996} seminal result that, in independent private value auctions, the auctioneer's benefit of using optimal vs.\ second-price auctions is smaller than the benefit of having just one additional bidder.} As such, in data-rich settings, the benefit of optimizing over general contracts may plausibly be outweighed by other (unmodeled) benefits of using simple, binary contracts.

Section~\ref{sec:sketch} below explains the idea behind Theorem~\ref{thm:main} and sheds light on the structure of the binary contracts that achieve the optimal convergence rate. These contracts take a simple form derived from a statistical test, where the principal uses a log-likelihood score to partition empirical signal frequencies into a ``pass'' (high wage) and ``fail'' (low wage) region.\footnote{As such, the high and low wage regions are described by systems of linear inequalities in $\mathbb{R}^{|A|-1}$, rather than by potentially more complicated sets in $\mathbb{R}^n$.} Importantly, we will see that to achieve the optimal convergence rate, the pass--fail cutoff must be chosen in the \emph{maximally lenient} way subject to (\ref{eq:IC}), in the sense of minimizing the agent's probability of failing the test conditional on choosing the target action.  As noted in the Introduction, such leniency may be in line with binary contracts observed in practice. Example~\ref{ex:finite-n} below highlights the importance of leniency in a parametric environment.

\medskip

{\noindent {\bf Suboptimal convergence under linear contracts.}} Section~\ref{sec:sketch} also illustrates that, unlike binary contracts, contracts where the wage varies finely with the signal realizations may approximate the first-best at a highly suboptimal rate. Below, we formalize this point for another class of frequently observed wage schemes, \textbf{\textit{linear contracts}}: Here, there exists $b_n: X\to[\underline w, \infty)$ such that $w(x^n)=\sum_{i=1}^n b_n(x_i)$ for all $x^n \in X^n$; that is, wages are linear in empirical frequencies of signals. If $X\subseteq\mathbb R$ (as in Example~\ref{ex:finite-n} below), this definition nests a common formulation of linear contracts in the literature: wage schemes $w$ that are linear in average signals. Let $C^{\text{lin}}_n(\mu, u,c, a^*)$ denote the second-best cost when, instead of optimizing over all IC and IR contracts, the principal is restricted to optimizing over IC and IR linear contracts.  

The following result shows that the convergence rate to the first-best under linear contracts is subexponential:

\begin{prop}\label{prop:linear} Suppose that $\inf_{v \in u([\underline w, \infty))} h''(v) > 0$.\footnote{This is a slight strengthening of the assumption that $\lim_{w \to \infty} u'(w) =0$ and holds, e.g., under CARA and CRRA (with $\underline w > 0$).} Under linear contracts, the second-best cost converges to the first-best subexponentially: There exists a constant $K>0$ such that
\begin{equation}\label{eq:linear convergence}
C^{{\rm lin}}_n(\mu, u,c, a^*)-C^{{\rm FB}}(u,c, a^*)\geq \frac{1}{n}K+o\left(\frac{1}{n}\right).
\end{equation}
Moreover, (\ref{eq:linear convergence}) holds with equality if $\mu_{a^*}\not\in {\rm co}\{\mu_a: a\not=a^*\}$ and $\underline w$ is sufficiently low.\footnote{Here ${\rm co}$ denotes the convex hull operator.}
\end{prop}

Note that, in the perfect monitoring limit where $n = \infty$, both binary and linear contracts can achieve the first-best cost (the latter claim requires the regularity assumption on $\mu$ in the ``moreover'' part of Proposition~\ref{prop:linear}).\footnote{The first-best is achieved by any $w$ such that $w(x^\infty) = h( c(a^*))$  for $x^\infty$ whose empirical frequency is $\mu_{a^*}$ and $w(x^\infty)\leq h(c(a))$ for $x^\infty$ whose empirical frequency is $\mu_{a}$ for some $a\not=a^*$. Clearly, such $w$ can always be made binary, and can be made linear under the regularity assumption in Proposition~\ref{prop:linear}.} However, Proposition~\ref{prop:linear} shows that, away from this limit, linear contracts are less effective at exploiting rich data than binary contracts, because they approximate the first-best at a much slower rate. Thus, regardless of the agent's utility and cost function and the principal's monitoring technology and target action, in the current setting the principal is always better off using binary contracts when the amount of data is rich enough: That is, there exists $N$ such that for all $n \geq N$,
$$C^{{\rm bin}}_n(\mu, u,c, a^*) < C^{{\rm lin}}_n(\mu, u,c, a^*).$$

\medskip

{\noindent {\bf Illustrative example.}} Before sketching the proof of Theorem~\ref{thm:main} and Proposition~\ref{prop:linear}, we illustrate how the convergence in these results plays out as a function of $n$. The following example considers a parametric environment where the performance of binary and linear contracts can be studied analytically at each $n$. Mitigating a potential concern that the predictions of the convergence rate  analysis might only apply at very large $n$, where both $C^{\rm bin}_n$ and $C^{\rm lin}_n$ may already be very close to $C^{\rm FB}$ and the performance gap between these two classes of contracts may be negligible, we show that lenient binary contracts start to come close to the first-best at fairly moderate $n$ and outperform linear contracts by a significant margin:

\begin{ex}\label{ex:finite-n} The agent has a binary action set $A = \{0, 1\}$, effort cost $c(a) = ca$ for some $c \in (0,1)$, and CARA utility function $u(w) = 1-\exp[-\eta w]$ for some $\eta>0$. The lower bound on wages is $\underline w\leq-\frac{1}{\eta}\ln 2$. Conditional on action choice $a$, the principal observes $n$ i.i.d.\ Gaussian signals $x_i = a + \varepsilon_i$, where $\varepsilon_i \sim {\cal N}(0, 1)$. The first-best cost of implementing $a^*=1$ is $C^{\rm FB}= h (c) = -\frac{1}{\eta}\ln (1-c)$. 

We study analytically how well $C^{\rm FB}$ is approximated as a function of $n$ under binary vs.\ linear contracts (see Appendix~\ref{app:illustration} for all derivations). First, consider binary contracts that specify a cutoff $\gamma$ and pay a high wage $w^+_n$ if the average signal $\bar x := \frac{1}{n} \sum_{i=1}^n x_i$ exceeds $\gamma$ and a low wage $w^-_n$ otherwise, where $w^+_n > w^-_n$ are pinned down by requiring (\ref{eq:IC})--(\ref{eq:IR}) to bind. For all  $n > 2 \ln \frac{2}{1-c}$, the difference between the optimal implementation cost of $a^*=1$ under such contracts and $C^{\rm FB}$ can be upper-bounded by
\begin{equation}\label{eq:bd-bin}
\frac{2c}{\eta}\frac{\exp[-\frac{n}{2}]}{1-c-2\exp[-\frac{n}{2}]}.
\end{equation}
This bound decays exponentially at rate $1/2$ which, in line with Theorem~\ref{thm:main}, equals the KL divergence between the signal distributions ${\cal N}(0, 1)$ and ${\cal N}(1, 1)$ under actions $0$ vs. $1$. Based on this, it is easy to see that binary contracts can come close to the first-best at moderate $n$ under natural parameter values: Figure~\ref{fig:binlin} illustrates this for a lenient binary contract whose cutoff $\gamma = 0.1$ is only slightly above the expected signal under the non-target action $0$.\footnote{We extend the domain of the figure to non-integer $n$ by treating $\overline x$ as a single signal draw that is distributed $\overline x \sim \mathcal{N} (a, \frac{1}{n})$ conditional on action $a$.} In line with the importance of leniency discussed above, convergence under this contract is substantially faster than under a symmetric binary contract with a more demanding cutoff $\gamma = 0.5$ that is exactly in between the expected signals under the two actions. 

\begin{figure}[t]
\begin{center}
\begin{overpic}[unit=1mm,scale=0.15]{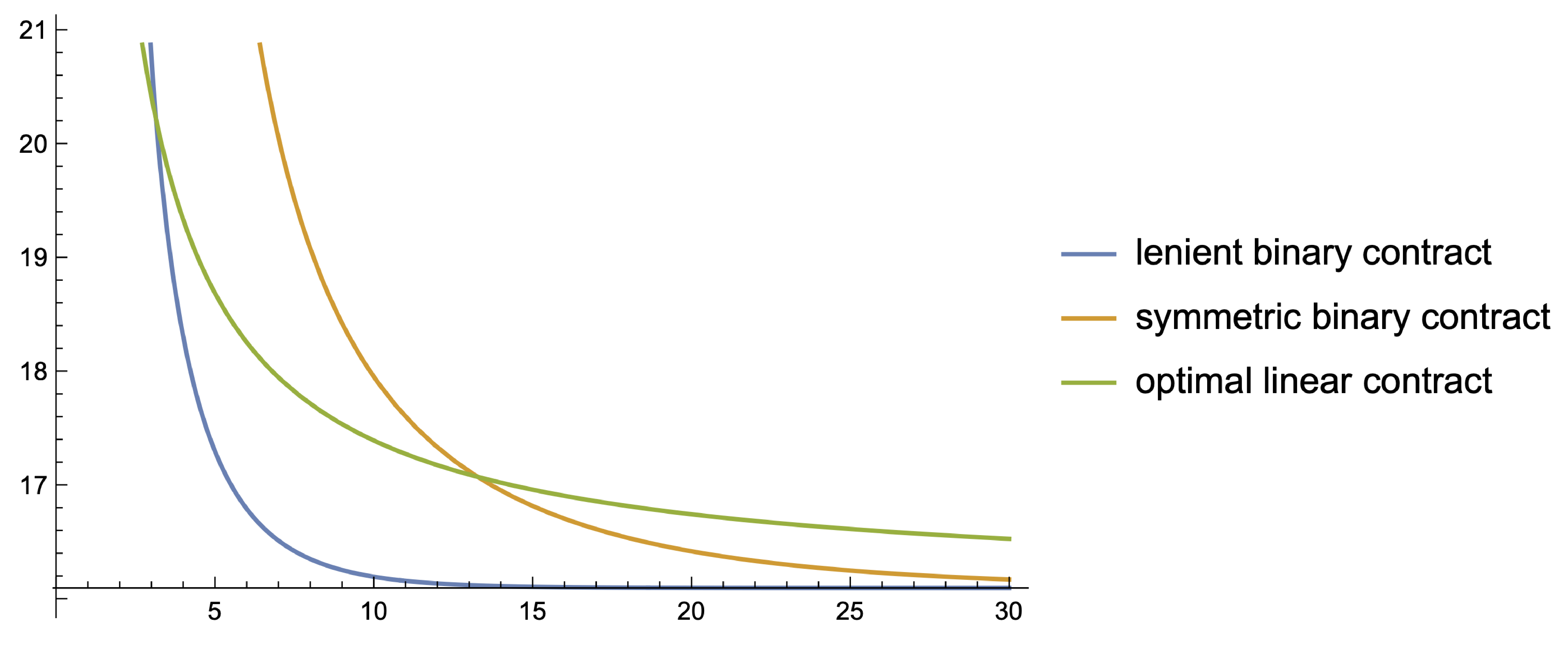}
\put(-12,39){{\tiny $\stackrel{\text{implementation}}{\text{cost}}$}}
\put(64,1){{\tiny $n$}}
\end{overpic}
\end{center}
\begin{caption}
{\label{fig:binlin} \footnotesize Implementation costs in Example~\ref{ex:finite-n}  as a function of $n$ when $c =0.8$, $\eta=0.1$, so $C^{\rm FB} \approx 16.1$. The lenient (resp.\ symmetric) binary contract uses a cutoff $\gamma=0.1$ (resp.\ $\gamma=0.5$).}
\end{caption}
\end{figure}

In contrast, consider the optimal IC-IR contract whose wage $w(x^n) = \alpha_n \overline x + \beta_n$ is linear in average signals $\overline x$.\footnote{Since signal distributions in this example are unbounded, this contract ignores the requirement that wages are bounded below. One way to avoid this issue would be to consider the modified contract given by $\tilde w(x^n):=\max\{w(x^n), \underline w\}$ for each $x^n$. Under this modification, the gap to $C^{\rm FB}$ remains of order $1/n$.} For all $n$, the gap between the implementation cost under this contract and $C^{\rm FB}$ can be shown to equal
\begin{equation}\label{eq:bd-lin}
\frac{1}{n} \frac{\left(\ln (1-c)\right)^2}{2\eta} .
\end{equation}
In line with Proposition~\ref{prop:linear}, (\ref{eq:bd-lin}) is only of order $1/n$, and thus tends to be significantly larger than the exponential bound (\ref{eq:bd-bin}) at moderate $n$. For example, in Figure~\ref{fig:binlin}, the optimal linear contract outperforms the lenient binary contract at very small $n$, but the ranking quickly reverses, and the gap in implementation costs remains significant throughout the domain of the figure. \finex

\end{ex}

%%%%%%%%%%%%%%%%%%%%%%%%%%%%%%%%%%%%%%%%%%%%%%%%%%
%%%%%%%%%%%%%%%%%%%%%%%%%%%%%%%%%%%%%%%%%%%%%%%%%%%

\subsection{Proof Sketch of Theorem~\ref{thm:main} and Proposition~\ref{prop:linear}}\label{sec:sketch}

For expositional simplicity, we focus on binary action sets, $A=\{0, 1\}$ with $a^*=1$. 

{\bf Step 1: Variance minimization.} Without loss of optimality, consider any uniformly bounded sequence of (utility) contracts $v_n$ with binding (\ref{eq:IR'}) (i.e., ${\mathbb E}_{1}[v_n(x^n)]=c(1)$). Then
\begin{equation}\label{eq:variance efficiency}
\underbrace{{\mathbb E}_{1}[h(v_n(x^n))]-C^{\text{FB}}(u,c, a^*)}_{\text{inefficiency}} = \underbrace{{\mathbb E}_{1}[h(v_n(x^n))]-h\left({\mathbb E}_{1}[v_n(x^n)]\right)}_{\text{Jensen-inequality gap}} 
 \approx {\rm Var}_{1}[v_n(x^n)],
\end{equation}
where ``$\approx$'' denotes convergence to $0$ at the same exponential rate. That is, the efficiency loss relative to the first-best is equal to the Jensen-inequality gap of the convex function $h$ with respect to the random variable $v_n(x^n)$; by Taylor expansion arguments \citep[e.g.,][]{liao2018}, the latter has the same exponential decay rate as the variance of $v_n(x^n)$. 
Thus, to obtain the optimal convergence rate to the first-best, the principal must choose $v$ to maximize the decay rate of the agent's utility variance ${\rm Var}_{1}[v_n(x^n)]$ subject to (\ref{eq:IC'}). This reflects the familiar tradeoff in moral hazard problems \citep[e.g.,][]{laffont2009} of seeking to make contracts as ``safe'' for the risk-averse agent as possible without violating incentive compatibility.

{\bf Step 2: Binary test contracts.} Next, we construct simple binary ``test'' contracts under which ${\rm Var}_1 [v_n(x^n)]$ decays at rate ${\rm KL} (\mu_0, \mu_1)$. Denote by
\[
L_n:=\frac{1}{n}\sum_{i=1}^n\ln\frac{d\mu_{1}}{d\mu_{0}}(x_i) \] 
the \textit{\textbf{log-likelihood score}} of the realized signal sequence $x^n$, defined as the average log-likelihood ratio of action 1 vs.\ 0.
 Note that the expected scores under both actions, ${\mathbb E}_{1}[L_n] >0$ and ${\mathbb E}_{0}[L_n] <0$, are independent of $n$. By standard arguments \citep[e.g.,][]{laffont2009}, it is without loss  to restrict to contracts $v_n$ that are nondecreasing functions of $L_n$. Consider a sequence of binary contracts of the form
\begin{equation}\label{eq:test contract}
v_n(x^n)=
\begin{cases}
v^+_n \text{ if } L_n\geq \gamma  \\
v^-_n \text{ if } L_n<\gamma.
\end{cases} 
\end{equation}
Here $\gamma$ is some threshold with $\gamma\in ({\mathbb E}_{0}[L_n], {\mathbb E}_{1}[L_n]))$, and the utility payments $v^+_n> v^-_n$ are pinned down by requiring (\ref{eq:IR'}) and (\ref{eq:IC'}) to bind, which can be ensured at all large enough $n$.\footnote{That is, 
  $v^+_n :=  \frac{c(1){\mathbb P}_{0}[L_n<\gamma]-c(0){\mathbb P}_{1}[L_n<\gamma]}{{\mathbb P}_{1}[L_n\geq\gamma]-{\mathbb P}_{0}[L_n\geq\gamma]}$ and
$v^-_n :=  \frac{c(0){\mathbb P}_{1}[L_n\geq\gamma]-c(1){\mathbb P}_{0}[L_n\geq\gamma]}{{\mathbb P}_{1}[L_n\geq\gamma]-{\mathbb P}_{0}[L_n\geq\gamma]}$, which converge to $c(1)$ and $c(0)$, respectively, as $n\to\infty$. By Assumption~\ref{asp}.2, $v^+_n$, $v^-_n \in u([\underline w, \infty))$ for large enough $n$.\label{fn:wages}} That is, the agent is rewarded with the higher utility $v^+_n$ whenever his score is above the threshold (i.e., he ``passes the test''), and he is punished with the lower utility $v_n^-$ if his score is below the threshold (i.e., he ``fails the test'').

Clearly, any such sequence approximates the first-best as $n \to \infty$, because the test becomes arbitrarily precise, i.e., $\lim_{n\to\infty}{\mathbb P}_{1}[L_n\geq \gamma]=1$ and $\lim_{n\to\infty}{\mathbb P}_{0}[L_n\geq \gamma]=0$, by the law of large numbers and the fact that $\gamma \in ({\mathbb E}_{0}[L_n], {\mathbb E}_{1}[L_n]))$.  However, we need to analyze which choice of the threshold $\gamma$ achieves the fastest convergence rate.

\begin{figure}[t]
\begin{center}
\begin{overpic}[unit=1mm,scale=.25]{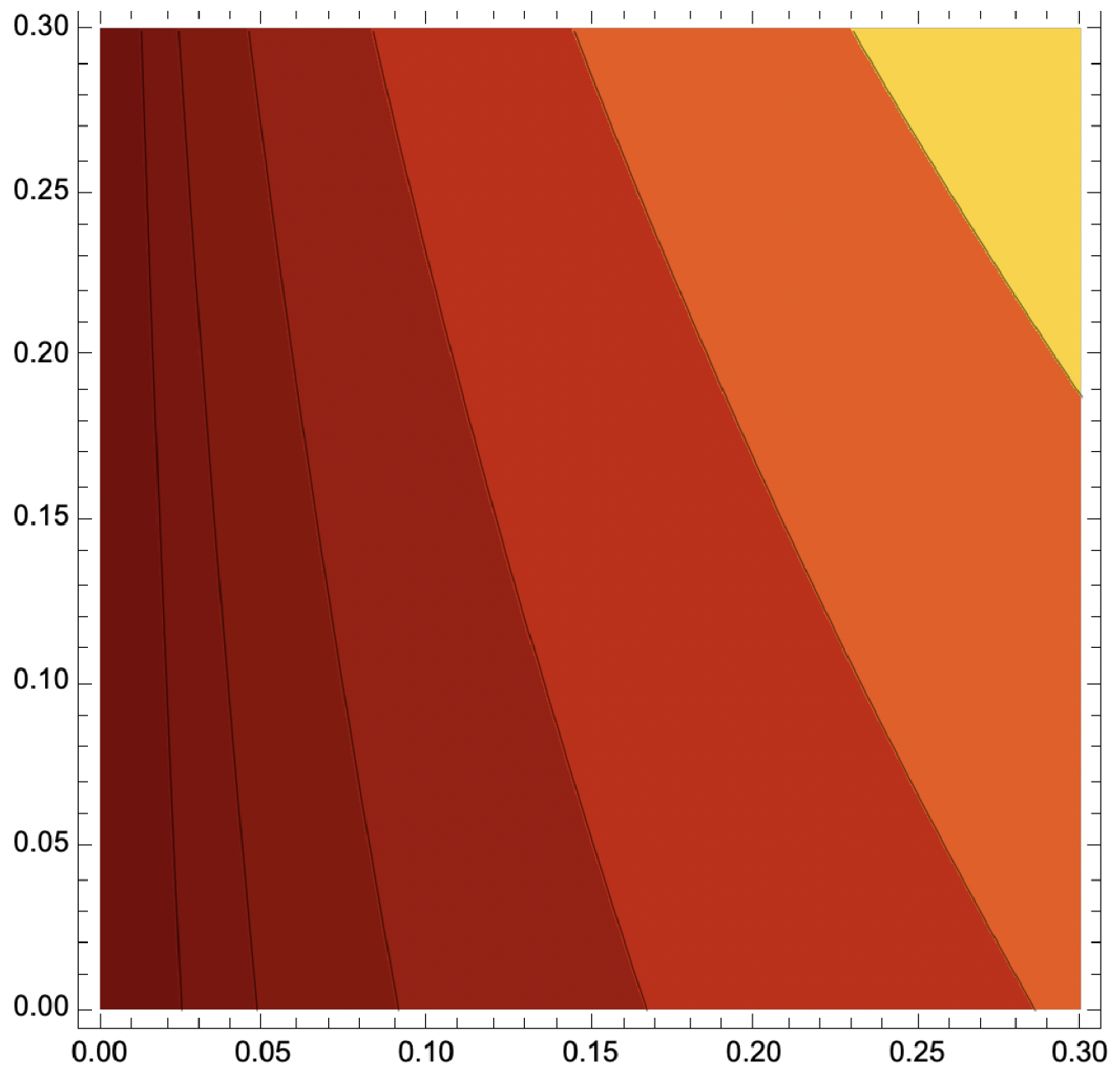}
\put(-25,90){{\tiny $\stackrel{\text{prob.\ of}}{\text{false positives}}$}}
\put(80,-5){{\tiny $\stackrel{\text{prob.\ of}}{\text{false negatives}}$}}
\end{overpic}
		\end{center}
		\begin{caption}
{\label{fig:tradeoff} \footnotesize Contour curves of $ {\rm Var}_{1}[v_n(x^n)]$ under contract (\ref{eq:test contract}). Given $\gamma$, $n$ determines the probabilities of false positives and false negatives, which jointly determine the utility variance. }
\end{caption}

\end{figure}

To this end, observe that for any fixed $n$, the optimal choice of $\gamma$ must trade-off reducing two mistakes: On the one hand, the probability of 
 \textit{\textbf{false positives}}, ${\mathbb P}_{0}[L_n\geq \gamma]$, is lower the \emph{higher} is $\gamma$; reducing this is relevant for (\ref{eq:IC'}), as it makes action 0 less attractive to the agent. On the other hand, the probability of \textit{\textbf{false negatives}}, ${\mathbb P}_{1}[L_n< \gamma]$, is lower the \emph{lower} is $\gamma$; reducing this is relevant for both (\ref{eq:IC'}) and (\ref{eq:IR'}), as it makes action 1 more attractive to the agent. Crucially, we show that as $n$ grows large, false negatives are the dominant force affecting the agent's utility variance, because
 \[
{\rm Var}_{1}[v_n (x^n)]\approx  {\mathbb P}_{1}[L_n< \gamma].
\]
Thus, as $n \to \infty$, to minimize utility variance, the principal should optimally choose a \textit{\textbf{maximally lenient}} threshold $\gamma \searrow {\mathbb E}_{0}[L_n]$}.\footnote{For simplicity, (\ref{eq:test contract}) considered sequences of contracts with a fixed threshold $\gamma$. In this case, the optimal decay rate of ${\rm Var}_{1}[v_n (x^n)]$ is approximated by choosing $\gamma \in ({\mathbb E}_{0}[L_n], {\mathbb E}_{1}[L_n])$ arbitrarily close to ${\mathbb E}_{0}[L_n]$. Exactly attaining the optimal decay rate instead requires using an appropriate sequence of thresholds $\gamma_n \in ({\mathbb E}_{0}[L_n], {\mathbb E}_{1}[L_n])$ with $\gamma_n \to {\mathbb E}_{0}[L_n]$.} 
To illustrate, Figure~\ref{fig:tradeoff} shows the contour curves of $ {\rm Var}_{1}[v(x^n)]$ as a function of the probabilities of false positives and false negatives. Observe that at large values of $ {\rm Var}_{1}[v_n(x^n)]$ (i.e., at small $n$), false positives and false negatives have a fairly symmetric effect on utility variance; however, as $ {\rm Var}_{1}[v_n(x^n)]$ approaches $0$ (i.e., at large $n$), the contour curves become arbitrarily steep, illustrating the dominant effect of false negatives in this region. Intuitively, as monitoring becomes more and more precise, (\ref{eq:IC'}) becomes less important than (\ref{eq:IR'}) (as captured by a vanishing shadow value in the principal's optimization problem). As a result, reducing false positives becomes relatively unimportant, as this only affects (\ref{eq:IC'}).

\begin{figure}[t]
\begin{center}
\includegraphics[scale=0.35]{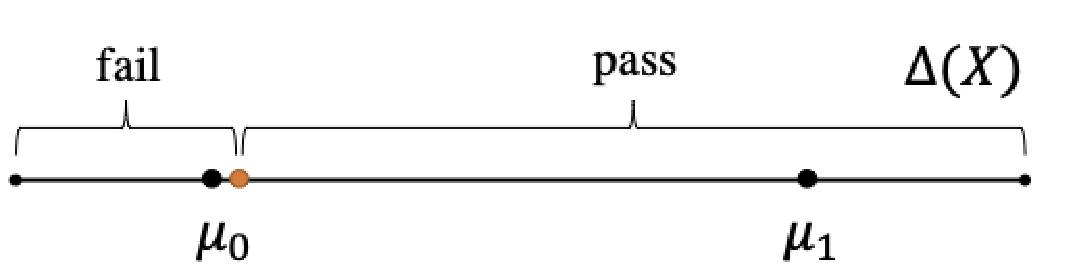}
\end{center}
\begin{caption}
{\label{fig:test KL} \footnotesize Contract (\ref{eq:test contract}) with binary signals. The space of empirical signal frequencies $\Delta (X)$ is divided into two regions, ``pass'' ($L_n\geq \gamma$) and ``fail'' ($L_n< \gamma$). As $\gamma$ approaches the maximally lenient threshold ${\mathbb E}_{0}[L_n]$, the cutoff between the two regions approaches $\mu_0$.}
\end{caption}
\end{figure}

Finally, we observe that as $\gamma$ approaches the maximally lenient threshold ${\mathbb E}_{0}[L_n]$, the probability of false negatives, and hence the utility variance, decays exponentially at rate ${\rm KL} (\mu_0, \mu_1)$. This observation essentially corresponds to Stein's lemma, a classical result in hypothesis testing \citep[e.g.,][]{cover1999}, and can be proved using Sanov's theorem from large deviation theory. Sanov's theorem states that for any set $D \subseteq \Delta(X)$ that is equal to the closure of its interior, the probability ${\mathbb P}_{1}[ \nu_n\in D]$ of observing an empirical frequency $\nu_n$ in $D$ conditional on action 1 decays exponentially at rate $\inf_{\nu \in D} {\rm KL} (\nu, \mu_1)$ (i.e., it decays faster the farther $D$ is from the theoretical signal distribution $\mu_1$ under action 1). Figure~\ref{fig:test KL} illustrates this in the case of binary signals: Applying Sanov's theorem to the  ``fail'' region $D$, the probability ${\mathbb P}_{1}[\nu_n \in D]$ of false negatives under the maximally lenient threshold decays at rate $\inf_{\nu \in D} {\rm KL} (\nu, \mu_1)= {\rm KL} (\mu_0, \mu_1)$.

{\bf Step 3: General contracts cannot do better.} It remains to show that general, non-binary contracts cannot converge to the first-best faster than at the exponential rate ${\rm KL}(\mu_0, \mu_1)$.  To illustrate the idea, consider sequences of contracts of the form
\[
v_n(x^n)=f(L_n),
\]
for some function $f$ that is continuously differentiable at ${\mathbb E}_1[L_n]$. For example, this includes linear contracts where the wage is a fixed affine function of the score $L_n$ at all $n$.\footnote{The logic extends to general sequences of linear contracts $w_n (x^n)=\sum_{i=1}^n b_n(x_i)$, where the functions $b_n : X \to [\underline w, \infty)$ can vary with $n$, as in Proposition~\ref{prop:linear}.}  If $f'({\mathbb E}_1[L_n])\not=0$ (i.e., there is some wage variation around ${\mathbb E}_1[L_n]$, as is the case for linear contracts), then the delta method implies that ${\rm Var}_{1}[v_n (x^n)]$ is of order $\frac{1}{n}f'({\mathbb E}_1[L_n])$, as $L_n$ is the sample average of $n$ i.i.d.\ draws of $\ln\frac{d\mu_{1}(x_i)}{d\mu_{0}(x_i)}$. Thus, the utility variance, and hence, by (\ref{eq:variance efficiency}), the efficiency loss relative to the first-best, vanishes subexponentially, i.e., more slowly than in Step 2.

Therefore, to achieve the optimal convergence rate, $f$ must be flat around ${\mathbb E}_1[L_n]$, as is the case for the above binary test contracts. Moreover, as we formalize in Appendix~\ref{app:lowerbd}, the convergence rate is higher the greater a flat region $f$ displays around ${\mathbb E}_1[L_n]$. At the same time, to satisfy (\ref{eq:IC'}), the flat region cannot extend below ${\mathbb E}_0 [L_n]$. Since the maximally lenient binary contract maximizes the flat region subject to this constraint, no other contract can outperform its convergence rate  ${\rm KL}(\mu_0, \mu_1)$.

{\bf Non-binary actions.} The above logic extends readily to non-binary action sets. Specifically, we use binary contracts of the following form: For each $a \in A^{-} (c, a^*)$, consider the log-likelihood score $L_n(a):= \frac{1}{n}\sum_{i=1}^n\ln\frac{d\mu_{a^*}}{d\mu_{a}}(x_i)$ and a threshold $\gamma (a)$. Award a high utility payment $v^+_n$ if $L_n (a) \geq \gamma(a)$ for \emph{all} $a \in A^{-} (c, a^*)$ and a low payment $v^-_n$ otherwise, where $v^+_n$ and $v^-_n$ are again pinned down by requiring (\ref{eq:IR'}) and (\ref{eq:IC'}) to bind. Here, the optimal rate of convergence is again achieved by setting each threshold $\gamma(a) \searrow \mathbb{E}_a [L_n (a) ]$ to be maximally lenient.

\begin{rem}[Contrast with ``shoot the agent'' contracts]\label{rem:mirrlees} In the setting where $n = 1$ and there is no lower bound on wages, \cite{mirrlees1999} shows that, even when the monitoring technology is highly imprecise, the first-best can (essentially) be achieved as long as signal likelihood ratios are unbounded. For this purpose, he uses binary contracts that severely punish the agent at signals with very low log-likelihood scores (i.e., he considers the limit where both the low wage and the score cutoff become arbitrarily negative). This argument does not apply in our setting at any $n$, as we imposed a lower bound $\underline w$ on wages. As a result, the binary contracts we constructed above are qualitatively quite different from Mirrlees' ``shoot the agent'' contracts. In particular, Mirrlees' contracts punish deviations extremely severely but with probability close to $0$; in contrast, our binary contracts punish deviations with a moderately low wage but with probability close to 1 as $n \to \infty$. Indeed, note that our binary contracts use a low wage $v^-_n \to c(0) > \underline w$ that remains bounded away from $\underline w$ at large $n$ (see footnote~\ref{fn:wages}).\footnote{In contrast, the lower bound $\underline w$ does bind at the general second-best contract at large $n$. The optimal convergence rate under binary contracts is not sensitive to the exact choice of $v^-_n$ (e.g., it can also be achieved by some contracts with $\gamma \searrow \mathbb{E}_0 [L_n]$ and $\lim v^-_n \in [\underline w, c(0))$, even though under such contracts (\ref{eq:IC'}) is slack).} Thus, even for arbitrarily low values of $\underline w$, our binary contracts are not an approximation of Mirrlees' construction. \finex
\end{rem}

\begin{rem}[Limit of optimal contracts]\label{rem:limit} While maximally lenient binary contracts achieve the optimal convergence rate to the first-best, the illustration above shows that the same is true for some other contracts: To converge at the optimal rate, we saw that wage schemes must become flat at likely signal realizations (i.e., empirical signal frequencies $\nu$ within distance ${\rm KL} (\mu_0, \mu_1)$ of $\mu_1$). But beyond maximally lenient binary contracts, this allows for contracts that display additional wage variation at very rare empirical signal frequencies (farther than distance ${\rm KL} (\mu_0, \mu_1)$ from $\mu_1$). 
 
However, even under the general second-best problem, the benefit to introducing such additional wage variation vanishes as $n \to \infty$: Any convergent sequence of optimal contracts has as its limit a maximally lenient binary contract.

To state this formally, again assume for simplicity that $A = \{0, 1\}$ with $a^* = 1$ (Appendix~\ref{sec:app-limit} extends the result to non-binary action sets). As noted, we can identify optimal contracts with functions $v^*_n$ of log-likelihood scores $L_n$.\footnote{More precisely, we treat $v^*_n$ as a function on the domain ${\rm co}\{\ln\frac{d\mu_{1}}{d\mu_{0}}(x) : x \in X \}$.} Endow the space of such functions with the topology of weak convergence (under the $L^2$-norm). 
\begin{prop}\label{prop:limit2}
For any weakly convergent sequence of optimal contracts $(v^*_n)$,
\[
\lim_{n\to\infty} v^*_{n} (L)=
\begin{cases}
c (1) &\text{ if } L> \gamma^*, \\
u(\underline w)  &\text{ if } L < \gamma^*, 
\end{cases}
\]
where $\gamma^* := \mathbb{E}_0 [L_n]$ is the maximally lenient threshold.\end{prop}

Proposition~\ref{prop:limit2} adds to the rationale for maximally lenient binary contracts in data-rich settings. We note that Proposition~\ref{prop:limit2} is neither implied by nor implies Theorem~\ref{thm:main}.\footnote{To see the latter point, note that Proposition~\ref{prop:limit2} is silent about the rate of convergence of $v^*_n$ to its binary limit contract. In particular, it does not rule out that this convergence is slower than the rate at which $v^*_n$ approximates the first-best.} \finex \end{rem}

\subsection{Ranking over Monitoring Technologies}\label{sec:ranking}

Theorem~\ref{thm:main} immediately implies the following ranking over monitoring technologies:

\begin{cor}\label{cor:ranking} Take any nonempty $A^{-} \subsetneq A$ and any two monitoring technologies $\mu$ and $\mu'$ with  $\min_{a \in A^{-}}{\rm KL} (\mu_a, \mu_{a^*})>\min_{a \in A^{-}}{\rm KL} (\mu'_a, \mu'_{a^*})$. Then, for any $u$ and $c$ with $A^{-} (c, a^*) = A^{-}$, there exists $N$ such that for all $n \geq N$,
\begin{center}
$C^{\rm{SB}}_n(\mu, u,c, a^*)\leq C^{{\rm bin}}_n(\mu, u,c, a^*)<C^{\rm{SB}}_n(\mu', u,c, a^*).$
\end{center}
\end{cor}

Holding fixed a target action $a^*$, Corollary~\ref{cor:ranking} yields a (generically complete) ranking over monitoring technologies that quantifies their value to the principal in data-rich settings: Whenever the amount of data $n$ is rich enough, monitoring technologies $\mu$ are more valuable to the principal the higher is the index $\min_{a \in A^{-} (c, a^*) } {\rm KL} (\mu_a, \mu_{a^*})$. This is because, by Theorem~\ref{thm:main}, this index captures the optimal convergence rate to the first-best under $\mu$. Moreover, since the optimal convergence rate is achieved using binary contracts, the principal is better off using a superior monitoring technology $\mu$ along with a binary contract than optimizing over general contracts under any (even slightly) lower-ranked monitoring technology $\mu'$. 

Notably, Corollary~\ref{cor:ranking} provides the principal with \emph{detail-free} guidance for selecting between monitoring technologies: The ranking is independent of the agent's utility function $u$ and depends on his cost function $c$ only through the set $A^{-} (c, a^*)$ of actions that are less costly than the target action.

The following example illustrates a qualitative implication of the ranking:

\begin{ex}[Precise bad vs.\ good news]\label{ex:asymmetry}
Assume binary actions and signals,  $A = X=\{0, 1\}$ with $a^* = 1$, $c(1) > c(0)$. Consider monitoring technologies $\mu$ and $\mu'$ with
\[
 \mu_0(0)=0.8, \; \mu_1(1)=0.99 ; \quad \quad \quad \mu'_0(0)=0.99, \; \mu'_1(1)=0.8.
\]
Under both $\mu$ and $\mu'$, signal 1 is ``good news'' (more indicative of the target action $a^* =1$) and signal $0$ is ``bad news'' (more indicative of the deviation $a = 0$). The only difference is that the error probabilities under the two actions are flipped across $\mu$ and $\mu'$: $\mu$ provides precise bad news and relatively less precise good news (i.e., $\frac{\mu_0 (0)}{\mu_1 (0)} > \frac{\mu_1 (1)}{\mu_0 (1)}$), while $\mu'$ provides precise good news and relatively less precise bad news.

Observe that ${\rm KL}(\mu_0, \mu_1) \approx 3.2 >{\rm KL}(\mu'_0, \mu'_1) \approx 1.5$. Thus, by Corollary~\ref{cor:ranking}, whenever data is sufficiently rich, then regardless of the agent's utility and cost functions, $\mu$ is strictly more valuable to the principal than $\mu'$. Intuitively, as we saw in Section~\ref{sec:sketch}, the principal seeks to design tests under which false negatives decay as fast as possible, and for this monitoring technologies that provide precise bad news about the agent's action are more valuable than ones that provide precise good news. \finex
\end{ex}

The ranking in Corollary~\ref{cor:ranking} relies on the principal observing rich data, which reduces the comparison of monitoring technologies to a comparison of the corresponding convergence rates to the first-best. If the principal only observes a single signal draw ($n =1$), then \cite{kim1995} and \cite{jewitt2007} show that requiring $\mu$ to be more valuable to the principal than $\mu'$ regardless of the agent's preferences yields an order over monitoring technologies that extends \cites{blackwell} order. However, their order is more conservative than our ranking:\footnote{Based on the $n = 1$ order, $\mu$ dominates $\mu'$ if, for all $a\in A^-(c, a^*)$,  the distribution of $\frac{d\mu_{a^*}}{d\mu_{a}}(x)$ under $\mu_{a}$ is a strict mean-preserving spread of the distribution of $\frac{d\mu'_{a^*}}{d\mu'_{a}}(x)$ under $\mu'_{a}$. (This is an adaptation to our setting of the condition in \cite{kim1995} and \cite{jewitt2007}, who studied a continuous-action setting assuming the first-order approach).  In this case, by Jensen's inequality, ${\rm KL}(\mu_a, \mu_{a^*}) > {\rm KL}(\mu'_a, \mu'_{a^*})$ holds for all $a\in A^-(c, a^*)$, so $\mu$ dominates $\mu'$ in the sense of Corollary~\ref{cor:ranking}.} For instance, the monitoring technologies in Example~\ref{ex:asymmetry} are incomparable based on the $n =1$ order.

As noted, Corollary~\ref{cor:ranking} can be viewed as an analog for incentive problems of \cites{moscarini2002} rich-data ranking over information structures in learning settings (see Section~\ref{sec:lit}). However, reflecting the difference between the value of information for incentive provision vs.\ learning, our ranking is different from their ranking: For instance, the two monitoring structures in Example~\ref{ex:asymmetry} are equally valuable in the learning setting of \cite{moscarini2002}.

\section{Discussion}\label{sec:discussion}

\subsection{Severe Limited Liability}\label{sec:limliab}

Our main model (Assumption~\ref{asp}.2) assumed that $c(A)\subseteq {\rm int}\left(u([\underline w, \infty))\right)$, which ensured that the (\ref{eq:IR'}) constraint could be made to bind at large enough $n$. In this section, we suppose this assumption is violated, i.e., limited liability is so severe that (\ref{eq:IR'}) does not bind. We show that binary contracts continue to achieve the optimal convergence rate to the first-best. However, the analysis highlights a tradeoff between rent extraction and risk aversion that was absent in the main setting.

For expositional simplicity, we focus on binary actions, $A=\{0, 1\}$ with $a^*=1$ and $c(0) < c(1)$; Appendix~\ref{app:severe} considers non-binary $A$. The assumption $c(A)\subseteq {\rm int}\left(u([\underline w, \infty))\right)$ implies $u(\underline w)<c(0)$. Departing from this, we impose the severe limited liability restriction that $u(\underline w)\geq c(0)$; thus, even if taking action $0$ leads the agent to be punished with the lowest wage $\underline w$, he prefers this to his outside option of $0$.  This, together with (\ref{eq:IC'}), implies that (\ref{eq:IR'}) does not bind at any $n$. Moreover, the first-best cost, i.e., the implementation cost of action 1 under perfect monitoring, is now
\[
C^{\rm{FB}}(u,c, a^*)=h \left(u(\underline w)+c(1)-c(0) \right).
\] 
We assume $u(\underline w)+c(1)-c(0)<\lim_{w\to\infty}u(w)$, so that $C^{\rm{FB}}(u,c, a^*)$ is well-defined. 

The following result shows that the optimal rate of convergence to the first-best can again be achieved using binary contracts. However, the rate of convergence is slower than in Theorem~\ref{thm:main}: Rather than being given by the KL divergence  ${\rm KL} (\mu_0, \mu_1)$, the convergence rate is now given by the Chernoff distance
\begin{equation}\label{eq:Chernoff}
{\rm Ch}(\mu_0, \mu_1) := \min_{\nu\in\Delta (X)} \max \{{\rm KL}(\nu, \mu_0), {\rm KL}(\nu, \mu_1)\}.
\end{equation}
Note that ${\rm Ch}(\mu_0, \mu_1)$ captures the KL divergence from distributions $\mu_0$ and $\mu_1$ to their KL-midpoint, as any minimizer $\nu$ in (\ref{eq:Chernoff}) must satisfy ${\rm KL} (\nu, \mu_0) = {\rm KL}(\nu, \mu_1)$. Thus, ${\rm Ch}(\mu_0, \mu_1)$ is smaller than ${\rm KL} (\mu_0, \mu_1)$ (and, unlike KL divergence, is symmetric).

\begin{thm}\label{thm:severe}  Assume $A = \{0, 1\}$ with $a^* =1$. Under both general and binary contracts, the second-best cost converges to the first-best exponentially at rate ${\rm Ch}(\mu_0, \mu_1)$:
\[
C^{\rm{SB}}_n(\mu, u,c, a^*)-C^{\rm{FB}}(u,c, a^*)=\exp[-{\rm Ch}(\mu_0, \mu_1) n+o(n)];
\]
\[
C^{{\rm bin}}_n(\mu,  u,c, a^*)-C^{\rm{FB}}(u,c, a^*)=\exp[-{\rm Ch}(\mu_0, \mu_1) n+o(n)].
\]
\end{thm}

The difference between Theorems~\ref{thm:main} and \ref{thm:severe} reflects that different binary contracts yield the optimal rate of convergence to the first-best in each setting. While the optimal convergence rate in Theorem~\ref{thm:main} was achieved by a binary test contract with a maximally lenient threshold $\gamma  \searrow \mathbb{E}_0 [L_n]$, in the current setting it is achieved by a binary test contract with a \emph{symmetric} threshold $\gamma = 0$. Indeed, as discussed in Section~\ref{sec:sketch}, using a maximally lenient threshold maximizes the rate at which false negatives decay, which was the dominant consideration in our main model at large $n$. In contrast, in the current setting, it turns out that both false positives and false negatives remain important sources of inefficiency even at large $n$. This necessitates using a symmetric binary test, as such a test equalizes the decay rates of false negatives and false positives, both of which equal ${\rm Ch} (\mu_0, \mu_1)$ \citep[e.g.,][]{cover1999}.

The equal concern for false positives and false negatives in the current setting reflects a tradeoff between risk aversion and rent extraction that was absent in the main model. To illustrate, we can decompose the difference between the implementation cost under any contract $v$ and the first-best cost as follows:
\begin{eqnarray*}
&&{\mathbb E}_1[h(v(x^n))]- C^{\rm{FB}}(u,c, a^*)\\
&=&\underbrace{{\mathbb E}_1[h(v(x^n))]-h({\mathbb E}_1[v(x^n)])}_{\text{Jensen-inequality gap}} \, + \, \underbrace{h({\mathbb E}_1[v(x^n)])-h(u(C^{\rm{FB}}(u,c, a^*)))}_{\text{information rent}}. 
\end{eqnarray*}
The first term is the Jensen-inequality gap under $v$, which reflects the agent's risk aversion. As in Section~\ref{sec:sketch}, this gap decays at the same rate as ${\rm Var}_1 [v(x^n)]$ as $n \to \infty$, which under binary contracts is governed by the decay rate of false negatives. The second term reflects that the agent's consumption utility ${\mathbb E}_1 [v(x^n)] = {\mathbb E}_0 [v(x^n)] + c(1) - c(0)$ (by (\ref{eq:IC'})) exceeds his utility $u(C^{\rm{FB}}(u,c, a^*)) = u(\underline w) + c(1) - c(0)$ under the first-best; that is, even at large $n$, the agent receives an information rent. Reducing the probability of false positives reduces ${\mathbb E}_0 [v(x^n)]$ and hence the size of this rent. The second term was absent in our main model, as there (\ref{eq:IR'}) could be made to bind at large $n$, giving the agent zero rent.

Notably, with non-binary actions, Appendix~\ref{app:severe} shows that the above tradeoff applies only to the least costly deviation $\hat a \in \argmin_A c$, while the logic for all other deviations is analogous to our main model. As a result, the optimal convergence rate is achieved by a binary contract that employs a mix of maximally lenient and symmetric tests.\footnote{Specifically, as in Section~\ref{sec:sketch}, this contract sets thresholds $\gamma (a)$ for each $a \in A^- (c, a^*)$, and pays a high wage iff the score $L_n (a)$ exceeds $\gamma (a)$ for all $a$. Here $\gamma(a) \searrow \mathbb{E}_ a [L_n (a)]$ is maximally lenient for all $a \in A^- (c, a^*)$, except for the least costly action $\hat a$, for which $\gamma(\hat a) = 0$ is the symmetric threshold.} Depending on which deviation is the dominant source of inefficiency at large $n$, the convergence rate is either $\min_{a \in A^- (c, a^*) \setminus \{\hat a\} } {\rm KL} (\mu_a, \mu_{a^*})$ as in Theorem~\ref{thm:main} or ${\rm Ch} (\mu_{\hat a}, \mu_{a^*})$ as in Theorem~\ref{thm:severe}. We note that, analogous to Proposition~\ref{prop:linear}, convergence under linear contracts is again subexponential.

\begin{rem}[Risk-neutral case]\label{rem:neutral} Under severe limited liability (unlike in our baseline model), there is a gap between the first-best and second-best at each $n$ even when $u$ is linear. As is well-known, the optimal contract in this setting (if it exists) is binary if $|A| = 2$, but otherwise is in general complicated \citep[e.g.,][]{dutting2019}. Moreover, even if $|A| = 2$, the optimal binary contract takes an extreme, ``minimally lenient'' form very different from the contracts used in Theorems~\ref{thm:main}--\ref{thm:severe}: It pays a high wage only if $x^n$ \emph{maximizes} the log-likelihood score $L_n$, an event whose probability vanishes as $n \to \infty$ even under the target action. \finex
\end{rem}

\subsection{More General Formulations of Rich Monitoring Data}\label{sec:general}

\subsubsection{Non-i.i.d.\ Signals}\label{sec:non-iidl}

While our main model assumed that the principal observes $n$ i.i.d.\ draws of signals from a fixed monitoring technology $\mu$, Theorem~\ref{thm:main} extends to settings where $n$ indexes more general sequences of increasingly precise monitoring technologies. 
 
 Appendix~\ref{app:non-iid} formalizes this by considering sequences $(\mu^n)$ of monitoring technologies on a signal space $Z$ with the following key property: Under each chosen action $a$, the distributions of the log-likelihood scores $L_n(a') = \frac{1}{n} \ln\frac{d\mu^n_{a^*}}{d\mu^n_{a'}}(z)$ ($a' \neq a^*$) concentrate on deterministic limits as $n \to \infty$, where these limits are distinct across different chosen actions $a$ and the convergence is described by a well-behaved rate function $I_a$ (in the sense of the large deviation principle). Beyond our main model, this setting nests other natural examples of rich or precise monitoring data, such as:

\smallskip
 
{\it Vanishing observation noise:} When the agent chooses action $a \in A \subseteq \mathbb{R}$, the principal observes a single signal $x = a + \kappa_n \varepsilon$, where $\varepsilon$ is a standard normal noise term and the scaling factor $\kappa_n \to 0$ as $n \to \infty$. 

{\it Serially correlated signals:} When the agent chooses action $a$, the principal observes $n$ signals $x_1, \ldots, x_n$. However, instead of being i.i.d., these signals follow a (well-behaved) Markov chain whose transition kernel depends on $a$. 

\smallskip

In this more general setting, Theorem~\ref{thm:general} in Appendix~\ref{app:non-iid} shows that maximally lenient binary contracts again achieve the optimal convergence rate to the first-best. The only novelty is that the KL-based convergence rate in Theorem~\ref{thm:main} is replaced by a more general expression based on the rate function $I_{a^*}$.

\subsubsection{Adjustable Actions}\label{sec:adjust}

Our main model assumed that the principal observes $n$ i.i.d.\ signals about a fixed, one-shot action $a$. However, Theorem~\ref{thm:main} generalizes to settings where the agent can repeatedly adjust his action, as long as adjustments are infrequent relative to the frequency of signals.\footnote{This may capture, for instance, the use of digital productivity monitoring software, which tracks employees' on-screen activity at extremely small time intervals.}

Appendix~\ref{app:adjust} formalizes this by assuming that the agent can adjust his action $T$ times, i.e., sequentially chooses actions $a_1, \ldots, a_T \in A$ at a total cost of $\frac{1}{T} \sum_{t=1}^Tc(a_t)$. Each action $a_t$ generates $\frac{n}{T}$ i.i.d.\ draws of signals from $\mu_{a_t}$ that are observed by both the principal and agent. The agent can condition his choice $a_t$ on all past signal realizations. To keep the departure from the main model minimal, we continue to assume that the principal chooses contracts that offer a (one-shot) payment $w(x^n)$ as a function of the entire sequence $x^n$ of $n$ signals, and that the principal's first-best payoff is attained by a constant action profile with $a_t = a^*$ for all $t$. To capture that action adjustments are infrequent relative to the number of signals, we consider the rate of convergence of the principal's second-best payoff to her first-best payoff in the data-rich limit as $n \to \infty$, while holding fixed $T$. 

Theorem~\ref{prop:adjusment} in Appendix~\ref{app:adjust} shows that this convergence rate is again the same under general and binary contracts, and is given by $\frac{1}{T}\min_{a \in A^-(c, a^*)}{\rm KL} (\mu_a, \mu_{a^*})$. Thus, inefficiency is greater the more frequently the agent can adjust his action. Similar to our main model, the optimal convergence rate is achieved by a binary contract that employs a sequence of maximally lenient tests (one for each possible deviation at each $t$) and requires the agent to pass all tests to receive the high wage. Notably, the optimal convergence rate is the same as in the setting where the agent chooses an action sequence $(a_1, \ldots, a_T)$ but does not observe the signals generated by past actions; the latter can be viewed as a special case of our main model with $T$-dimensional actions and signals. Thus, whether or not the agent can condition his action choices on past data has a negligible effect on the principal's payoffs as the amount of data grows rich.

\subsection{Concluding Remarks}\label{sec:conclusion}

To offer a new perspective on the use of simple but suboptimal contracts, we studied moral hazard problems where the principal has access to rich monitoring data. In this setting, we highlighted that the convergence rate of the principal's payoffs to the first-best as data grows rich is a useful measure to quantify and compare the performance of different classes of contracts. We found that a particular class of widely observed contracts---binary wage schemes---that are simple in the sense of featuring the coarsest possible wage variation, is enough to achieve the optimal convergence rate. In contrast, linear contracts, another common compensation scheme that requires wages to vary finely with observed data (albeit in a simple manner), approximate the first-best at a highly suboptimal rate.

An important assumption for our analysis was that the agent's action space is discrete (in the sense of ${\rm KL} (\mu_a, \mu_{a^*})$ being bounded away from $0$). With this, we intend to capture settings where monitoring data is rich enough that it can effectively detect even relatively small deviations from the target action $a^*$: In particular, as $n$ grows large, the principal can design statistical tests that detect deviations with exponentially small errors. Our analysis applies for any arbitrarily small value of $\min_{a \in A^- (c,a^*)} {\rm KL} (\mu_a, \mu_{a^*}) > 0$, but more data $n$ is needed to design such accurate tests the smaller is $\min_{a \in A^- (c,a^*)} {\rm KL} (\mu_a, \mu_{a^*})$, leading to a slower convergence rate. In the extreme case where $A$ is continuous (and $\mu_a$ is appropriately continuous in $a$), the convergence to the first-best would be subexponential under both optimal and binary contracts. An analogous remark applies to the adjustable action setting in Section~\ref{sec:adjust}, which relied on signal arrivals being frequent relative to action adjustments (i.e., letting $n \to \infty$ while fixing $T$). If instead action adjustments and signals are equally frequent (i.e., $n = T$), then the impact of a single deviation becomes too small as $n \to \infty$ to be detected effectively by a statistical test. The analysis of the continuous action or frequent action adjustment settings would require a different approach from the current paper, as the large-deviation tools we relied on no longer apply. Thus, we leave this as a direction for future work.

\appendix

\section{Appendix}

\subsection{Preliminaries}

Throughout the Appendix, we denote by $L_n (a) := \frac{1}{n}\sum_{i=1}^n\ln\frac{d\mu_{a^*}}{d\mu_{a}}(x_i)$ the log-likelihood score for each $a \in A \setminus \{a^*\}$ and by $L_n := (L_n (a))_{a \neq a^*}$ the vector of log-likelihood scores.

\subsubsection{Cram\'er's Theorem and Other Statistical Preliminaries}

For each $a\in A$, $a' \in A \setminus \{a^*\}$ and $\ell\in\mathbb R$, define 
\[
I_{a, a'}(\ell):=\sup_{\lambda\in\mathbb R} \left( \lambda  \ell-\ln  \int \left(\frac{d\mu_{a^*}}{d\mu_{a'}}(x)\right)^{\lambda}d\mu_a(x) \right).
\] 
By Cram\'er's theorem \citep[][Theorem 2.2.3]{dembo2009}, function $I_{a, a'}$ describes the rate at which, under action $a$, the distribution of the log-likelihood score $L_n (a')$ concentrates on its expectation $\hat L_a(a'):={\mathbb E}_a[ L_n(a')]$ as $n \to \infty$: Specifically, $I_{a, a'}$ is continuous (on its effective domain), convex, and uniquely minimized at $\hat L_a (a')$ with $I_{a,a'}(\hat L_a(a'))=0$, and for any measurable set $B\subseteq\mathbb R$,
\begin{equation}\label{eq:cramer}
\begin{split}
- &\inf_{\ell\in {\rm cl} B}I_{a, a'}(\ell) \geq\limsup_{n\to\infty} \frac{1}{n}\ln  {\mathbb P}_a[ L_n(a')\in B]  \\
\geq
&\liminf_{n\to\infty} \frac{1}{n}\ln  {\mathbb P}_a[ L_n(a')\in B] \geq - \inf_{\ell\in {\rm int} B}I_{a,a'}(\ell).
\end{split}
\end{equation}
For each $\ell\in\mathbb R$,  Lemma 6.2.3(f) in \cite{dupuis2011} implies that
\begin{equation}\label{eq:variational}
I_{a^*,a}(\ell)= \inf_{\nu\in\Delta(X)}{\rm KL}(\nu, \mu_{a^*}) \, \text{ s.t. } \,  \int \ln\frac{d\mu_{a^*}}{d\mu_{a}}(x)d\nu(x) =\ell.
\end{equation}

The following lemma characterizes the two statistical distances used in the paper:

\begin{lem}\label{lem:KL}
For all $a \in A \setminus \{a^*\}$, we have the following two equalities:
\begin{equation}\label{eq:KL-lem}
{\rm KL}(\mu_{a}, \mu_{a^*})=\min_{\nu\in\Delta(X)} {\rm KL}(\nu, \mu_{a^*}) \text{ s.t. } \int \ln\frac{d\mu_{a^*}}{d\mu_{a}}(x)d\nu(x)\leq {\mathbb E}_{a}[ L_n(a)];
\end{equation}
\begin{equation}\label{eq:Ch-lem}
\begin{split}
{\rm Ch}(\mu_{a}, \mu_{a^*})=&\min_{\nu\in\Delta(X)} {\rm KL}(\nu, \mu_{a^*}) \text{ s.t. } \int \ln\frac{d\mu_{a^*}}{d\mu_{a}}(x)d\nu(x)\leq  0 \\
=& \min_{\nu\in\Delta(X)} {\rm KL}(\nu, \mu_{a}) \text{ s.t. } \int \ln\frac{d\mu_{a^*}}{d\mu_{a}}(x)d\nu(x)\geq  0.
\end{split}
\end{equation}
\end{lem}
\begin{proof}
To show (\ref{eq:KL-lem}), take any $\nu\in\Delta(X)$. Since $ \int\ln\frac{d\mu_{a^*}}{d\mu_{a}}(x)d\nu(x)={\rm KL}(\nu, \mu_a)-{\rm KL}(\nu, \mu_{a^*})$ and ${\mathbb E}_{a}[ L_n(a)] = \int \ln\frac{d\mu_{a^*}}{d\mu_{a}}(x)d\mu_a(x)=-{\rm KL}(\mu_a, \mu_{a^*})$, the constraint can be written as ${\rm KL}(\nu, \mu_a)+{\rm KL}(\mu_a, \mu_{a^*})\leq {\rm KL}(\nu, \mu_{a^*})$. By Gibbs' inequality, this implies that the minimum is achieved at $\nu=\mu_a$. 
Equation (\ref{eq:Ch-lem}) follows from standard information-theoretic arguments \citep[e.g.,][Section 11.9]{cover1999}.  \end{proof}

\subsubsection{Optimal Contracts}\label{app:optimal}

We briefly note some features of optimal contracts. First, for all large enough $n$, an optimal contract, i.e., a solution to the second-best problem (\ref{eq:SB}), exists. To see this, note that for each large enough $n$, there is a contract under which (\ref{eq:IC'}) and (\ref{eq:IR'}) hold with strict inequality; for example, such a contract can be obtained by adding a positive constant to the binary test contract $v_n$ defined in the proof of Theorem~\ref{thm:main} below. Given this, the existence of an optimal contract $v^*_n$ at all large enough $n$ follows from Section 4 in \cite{ke2023}. By standard arguments, $v^*_n$ can be written as a weakly increasing function of log-likelihood vectors $L_n \in \mathbb{R}^{A\setminus \{a^*\}}$.

Second, given the existence of contracts where (\ref{eq:IC'}) and (\ref{eq:IR'}) hold strictly, standard arguments \citep[e.g.,][Section 8.3]{luenberger1997} imply that the optimal contracts $v^*_n$ satisfy the Kuhn-Tucker conditions
\begin{equation}\label{eq:KKT'}
\begin{split}
h'(v^*_n( L_n))&=\lambda_n+\sum_{a \in A \setminus \{a^*\}}\kappa_n(a) \left(1-\exp[-n L_n(a)]\right)  \;\; \text{ if } v^*_n( L_n)>u(\underline w)\\
0 &\geq \lambda_n+\sum_{a \in A \setminus \{a^*\}}\kappa_n(a) \left(1-\exp[-n L_n(a)]\right)   \;\; \text{ if } v^*_n( L_n)=u(\underline w),
\end{split}
\end{equation}
where $\lambda_n\geq 0$ is the multiplier for (\ref{eq:IR'}) and $\kappa_n(a)\geq 0$ is the multiplier for the (\ref{eq:IC'}) constraint with respect to deviation $a\not=a^*$. This implies
\begin{equation}\label{eq:v*}
v^*_n( L_n)=\max \left\{ (h')^{-1}\left(\lambda_n+\sum_{a \in A \setminus \{a^*\}}\kappa_n(a) \left(1-\exp[-n L_n(a)]\right)\right), u(\underline w) \right\}.
\end{equation}

Finally, we note that the sequence $(v^*_n)$ is uniformly bounded. To see this, it is sufficient by (\ref{eq:v*}) to verify that $\limsup_{n\to\infty}\lambda_n+\sum_{a\not=a^*}\kappa_n(a)<\infty$. Suppose for a contradiction that $\lim_{k\to\infty}\lambda_{n_k}+\sum_{a\not=a^*}\kappa_{n_k}(a)=\infty$ for some subsequence $(n_k)$. Then for every strictly positive $L \in \mathbb{R}^{A\setminus \{a^*\}}$ (i.e., with strictly positive entries), (\ref{eq:KKT'}) implies $\lim_{k\to\infty}h'(v^*_{n_k}(L))=\infty$ and thus $\lim_{k\to\infty}h(v^*_{n_k}(L))=\infty$. By the weak law of large numbers, $ L_n$ converges in probability to ${\mathbb E}_{a^*}[ L_n]$, which is strictly positive. Thus, $\limsup_{n\to\infty}{\mathbb E}_{a^*}[h(v^*_n( L_n))]=\infty$, which contradicts optimality.

\subsection{Proof of Theorem~\ref{thm:main}}

Henceforth, let $A^-(c, a^*):=\{a\in A: c(a)<c(a^*) \}$, as in the main text, and $A^+(c, a^*):=\{a\in A: c(a)>c(a^*) \}$. We omit the dependency of $A^{-}$ and $A^+$ on $c$ and $a^*$ when there is no risk of confusion. By Assumption~\ref{asp}.3, $A=A^-\cup A^+\cup\{a^*\}$.

\subsubsection{Upper Bound on Inefficiency under Binary Contracts}

We first show that, as $n\to\infty$,  $C^{\rm bin}_n(\mu, u, c, a^*)-C^{\text{FB}}(u, c, a^*)$ vanishes at least as fast as at the exponential rate $\min_{a\in A^-}{\rm KL} (\mu_a, \mu_{a^*})$.

For each $a \in A^-$, fix some $\gamma(a) \in \left({\mathbb E}_{a} [ L_n(a)],{\mathbb E}_{a^*}[ L_n(a)]\right)$. For each $a'\in \{a^*\}\cup A^-$, define $p_{a', n}:={\mathbb P}_{a'}[ L_n(a)\geq \gamma(a), \,  \forall a\in A^-]$. Observe that, by the weak law of large numbers and the choice of $\gamma(a)$, we have
\[
\lim_{n\to\infty}p_{a^*, n}=1 \quad  \text{ and } \quad \lim_{n\to\infty}p_{a, n}=0  \text{ for all } a \in A^- .
\]
Consider the sequence of binary contracts $(v_n)$ given by
\[
v_n (x^n)= \begin{cases}
v^+_n &\text{ if }   L_n(a)\geq \gamma(a)   \text{ for all  } a\in A^- \\
v^-_n  &\text{ otherwise,} 
\end{cases}
\]
where $v^+_n$ and $v^-_n$ are defined by
\begin{align*}
v^+_n &:= \frac{(1-\max_{a\in A^-} p_{a, n})c(a^*)-(1-p_{a^*, n})\min_{a\in A^-}c(a)}{p_{a^*, n}-\max_{a\in A^-}p_{a, n}}, \\
 v_n^- &:=\frac{p_{a^*, n}\min_{a\in A^-}c(a)-\max_{a\in A^-}p_{a, n}c(a^*) }{p_{a^*, n}-\max_{a\in A^-}p_{a, n}}.
\end{align*}
Note that, up to restricting to large enough $n$, these contracts are well-defined (i.e., $v^+_n, v^-_n \in u([\underline w, \infty))$), because $\lim_{n\to\infty} v^+_n=c(a^*)$, $\lim_{n\to\infty} v^-_n=\min_{a\in A^-}c(a)$ and $c(A) \subseteq {\rm int} u([\underline w, \infty))$ by Assumption~\ref{asp}.2. Moreover, the choice of $v^+_n$ and $v^-_n$ implies 
\begin{align*}
&p_{a^*, n}v_n^++(1-p_{a^*, n})v_n^--c(a^*)=0 \quad \text{ and }\\
&p_{a^*, n}v_n^++(1-p_{a^*, n})v_n^--c(a^*)=\max_{a\in A^-}p_{a, n}v_n^++(1-\max_{a\in A^-}p_{a, n})v_n^--\min_{a\in A^-}c(a).
\end{align*}
The first line ensures that (\ref{eq:IR'}) holds with equality, as the LHS is the agent's expected payoff under $v_n$ to choosing action $a^*$. The second line ensures that choosing any $a \in A^-$ is (weakly) suboptimal for the agent. Moreover, choosing any $a\in A^+$ yields expected payoff strictly less than 0 for all large enough $n$, as $\lim_{n\to\infty}v^+_n=c(a^*)<c(a)$. Thus, $v_n$ also satisfies (\ref{eq:IC'}) for all large enough $n$.

Observe that 
\begin{equation}\label{eq:bin-cramer}
\begin{split}
\lim_{n\to\infty}\frac{1}{n}\ln(1-p_{a^*, n}) &= \lim_{n\to\infty}\frac{1}{n}\ln\left({\mathbb P}_{a^*}[ \exists a \in A^- \text{ s.t. }L_n(a)<\gamma(a)]\right) \\
&\leq -\min_{a\in A^-} \inf_{\ell\leq\gamma(a)}I_{a^*, a}(\ell),
\end{split}
\end{equation}
where the inequality holds by Cram\'er's theorem (\ref{eq:cramer}). Thus, for all large enough $n$,
\begin{eqnarray*}
&& C^{{\rm bin}}_n(\mu, u, c, a^*)-C^{\text{FB}}(u, c, a^*) \leq \mathbb E_{a^*}[h(v_n(x^n))]-h(c(a^*)) \\
 &=&p_{a^*, n} \left(h\left(v_n^+\right)-h (c(a^*))\right)+(1-p_{a^*, n})  \left(h\left(v_n^-\right)- h (c(a^*)) \right)\\
 &\leq&p_{a^*, n} \left(h\left(v_n^+\right)-h (c(a^*))\right) \leq \exp[- n \min_{a\in A^-} \inf_{\ell\leq\gamma(a)}I_{a^*, a}(\ell) +o(n)].
\end{eqnarray*}
Here, the penultimate inequality uses $\lim_{n\to\infty} v^-_n=\min_{a\in A^-}c(a) < c(a^*)$, and the last inequality uses (\ref{eq:bin-cramer}) and the fact that $v_n^+-c(a^*)=\frac{(1-p_{a^*, n})(\min_{a\in A^-}c(a)-c(a^*))}{p_{a^*, n}-\max_{a\in A^-}p_{a, n}}$ vanishes at the same exponential rate as  $1-p_{a^*, n}$.

Finally, for each $a\in A^-$, we have
\[
\sup_{ \gamma >{\mathbb E}_{a}[ L_n(a)]} \inf_{ \ell \leq\gamma}I_{a^*, a}(\ell)= \inf_{\ell\leq {\mathbb E}_{a}[ L_n(a)]}I_{a^*,a}(\ell) ={\rm KL}(\mu_a, \mu_{a^*}),
\]
where the second equality holds by (\ref{eq:variational})--(\ref{eq:KL-lem}) and the continuity of $I_{a^*,a}$. Thus, by choosing $\gamma(a)$ to be arbitrarily close to the maximally lenient threshold ${\mathbb E}_{a}[ L_n(a)]$ for each $a\in A^-$, it follows that  $C^{{\rm bin}}_n(\mu, u, c, a^*)-C^{\text{FB}}(u, c, a^*)$ vanishes at least as fast as at the exponential rate $\min_{a\in A^-}{\rm KL}(\mu_a, \mu_{a^*})$.

\subsubsection{Lower Bound on Inefficiency under General Contracts}\label{app:lowerbd}

Second, to complete the proof of Theorem~\ref{thm:main}, we show that, as $n\to\infty$, $C^{\rm{SB}}_n(\mu, u,c, a^*)-C^{\text{FB}}(u, c, a^*)$ vanishes no faster than at the exponential rate $\min_{a\in A^-}{\rm KL} (\mu_a, \mu_{a^*})$.

Pick any action $\hat a \in \argmin_{a\in A^-}{\rm KL}(\mu_a,\mu_{a^*})$.  Consider the relaxed problem where the agent's action set is restricted to $\{a^*, \hat a\}$. Below, we show that the difference between the principal's value under this relaxed problem and $C^{\text{FB}}(u,c, a^*)$ cannot vanish faster than at the exponential rate ${\rm KL}(\mu_{\hat a}, \mu_{a^*}) = \min_{a\in A^-}{\rm KL}(\mu_a,\mu_{a^*})$. This yields the desired conclusion, because under the original problem, $C^{\rm{SB}}_n(\mu, u,c, a^*)-C^{\text{FB}}(u,c, a^*)$ must vanish at a weakly slower rate than under the relaxed problem.

Consider any sequence of contracts $(v_n)$ under the relaxed problem, where (without loss of optimality) we write each $v_n$ as a weakly increasing function of $L_n :=L_n(\hat a)$.  
Let $\hat L_a:={\mathbb E}_a[L_n]$ for each $a \in \{a^*,\hat a\}$. 

We first show that in order for the agent's utility variance to decay (at least) exponentially, the contracts $v_n$ must become asymptotically flat on some interval around $\hat L_{a^*}$, where this interval must be larger the greater the rate of decay:

\begin{lem}\label{lem:flat}Consider any sequence of contracts $(v_n)$ such that ${\rm Var}_{a^*}[v_n(L_n)] \leq\exp[-\lambda n + o(n)]$ for some $\lambda>0$.  
Then for any interval $[\underline\gamma,\overline\gamma]\ni \hat L_{a^*}$ with $\inf_{\gamma<\underline\gamma} I_{a^*, \hat a}(\gamma) <\lambda$ and $\inf_{\gamma>\overline\gamma} I_{a^*,\hat a}(\gamma) <\lambda$,
\[
\lim_{n} \left( v_n(\overline \gamma)-v_n(\underline\gamma) \right)=0.
\]
\end{lem}

\begin{proof} Suppose for a contradiction that the result fails. Then, since each $v_n$ is weakly increasing, there is a subsequence $(n_k)$ such that $\lim_{k} \left( v_{n_k}(\overline \gamma)-v_{n_k}(\underline \gamma)\right)>0$. Thus, writing $m_{n_k}:={\mathbb E}_{a^*}[ v_{n_k}(L_{n_k})]$, we have
\begin{eqnarray*}
&&\liminf_{k\to\infty}\frac{1}{n_k}\ln{\rm Var}_{a^*}[v_{n_k}(L_{n_k})]=\liminf_{k\to\infty}\frac{1}{n_k}\ln {\mathbb E}_{a^*}[ \left(v_{n_k}(L_{n_k})-m_{n_k} \right)^2] \\
&\geq&\liminf_{k\to\infty}\frac{1}{n_k}\ln \left(\min \left\{{\mathbb P}_{a^*}[L_{n_k}\leq \underline \gamma], {\mathbb P}_{a^*}[L_{n_k}\geq \overline \gamma]\right\} \left(\frac{v_{n_k}(\overline \gamma)-v_{n_k}(\underline\gamma)}{2}\right)^2  \right) \\
&\geq& -\max\left \{\inf_{\gamma<\underline\gamma} I_{a^*,\hat a}(\gamma), \inf_{\gamma>\overline\gamma} I_{a^*, \hat a}(\gamma) \right\}>-\lambda,
\end{eqnarray*}
where the second inequality uses Cram\'er's theorem (\ref{eq:cramer}). This contradicts the assumption that ${\rm Var}_{a^*}[v_{n_k}(L_{n_k})] \leq \exp[-\lambda n_k+o(n_k)]$. 
\end{proof}

 Next, we show that under (\ref{eq:IC'}) the agent's utility variance cannot vanish faster than at the exponential rate ${\rm KL} (\mu_{\hat a}, \mu_{a^*})$:

 \begin{lem}\label{lem:IC bound}
 Consider any uniformly bounded sequence $(v_n)$ of contracts satisfying (\ref{eq:IC'}) under the relaxed problem. Then ${\rm Var}_{a^*}[v_n(L_n)] \geq \exp[-{\rm KL} (\mu_{\hat a}, \mu_{a^*}) n+o(n)]$.
  \end{lem}

\begin{proof}
Suppose otherwise. Then there exists $\lambda > {\rm KL} (\mu_{\hat a}, \mu_{a^*})$ such that ${\rm Var}_{a^*}[v_n(L_n)] \leq \exp[-\lambda n+o(n)]$. By (\ref{eq:variational})--(\ref{eq:KL-lem}), ${\rm KL} (\mu_{\hat a}, \mu_{a^*}) = I_{a^*,\hat a}(\hat L_{\hat a})$,  so $\lambda > I_{a^*,\hat a}(\hat L_{\hat a})$. Thus, by continuity of $I_{a^*, \hat a}$, there is $\underline\gamma<\hat L_{\hat a}$ with $\inf_{\gamma<\underline\gamma}I_{a^*, \hat a}(\gamma)<\lambda$. Moreover, by continuity of $I_{a^*, \hat a}$ and the fact that $I_{a^*, \hat a}(\hat L_{a^*})=0$, there is  $\overline\gamma>\hat L_{a^*}$ with $\inf_{\gamma>\overline\gamma}I_{a^*, \hat a}(\gamma)<\lambda$. By Lemma~\ref{lem:flat}, $\lim_{n\to \infty} \left(v_n(\overline\gamma)-v_n(\underline\gamma)\right)=0.$

At the same time, for each $a \in \{a^*,\hat a\}$, the fact that $\hat L_{a}$ is in the interior of $[\underline\gamma,\overline\gamma]$ and the weak law of large numbers implies that $\lim_{n\to\infty}{\mathbb P}_a \left[L_n\in [\underline\gamma,\overline\gamma]\right]=1$.
 Since the sequence of contracts is bounded, this implies $\lim_{n\to\infty} \left({\mathbb E}_{a^*}[v_n(L_n)]-{\mathbb E}_{\hat a}[v_n(L_n)]\right)=0$. But then $v_n$ violates (\ref{eq:IC'}) for all large enough $n$, a contradiction.
\end{proof}

Finally, consider a sequence $(v^*_n)$ of optimal contracts under the relaxed problem. By Appendix~\ref{app:optimal}, such a sequence exists (up to restricting to large $n$) and is uniformly bounded. The difference between the principal's cost and the first-best is
\begin{eqnarray*}
{\mathbb E}_{a^*}[h(v^*_n(L_n))]-h(c(a^*)) &=&{\mathbb E}_{a^*}[h(v^*_n(L_n))]-h({\mathbb E}_{a^*}[v^*_n(L_n)])+h({\mathbb E}_{a^*}[v^*_n(L_n)])-h(c(a^*)) \\
&\geq & {\mathbb E}_{a^*}[h(v^*_n(L_n))]-h({\mathbb E}_{a^*}[v^*_n(L_n)]) \\
&\geq &  {\rm Var}_{a^*}[v^*_n (L_n)] \frac{\inf_{L_n} h''(v_n(L_n))}{2},
\end{eqnarray*}
where the first inequality holds because ${\mathbb E}_{a^*}[v^*_n(L_n)]\geq c(a^*)$ by (\ref{eq:IR'}) and the second inequality uses the variance-based estimate of the Jensen-inequality gap \citep{liao2018}. Here $\inf_{L_n} h''(v_n(L_n))>0$, where the infimum is taken over all possible realizations of $L_n$, is bounded away from zero uniformly in $n$.  This follows by observing that (i) the contract wages are uniformly bounded, by the Kuhn-Tucker conditions (\ref{eq:KKT'}) and the assumption that $\lim_{w\to\infty}u'(w)=0$, and (ii) $u$ is $C^2$. 
Thus, by Lemma~\ref{lem:IC bound}, the difference relative to the first-best cannot vanish faster than at the exponential rate ${\rm KL}(\mu_{\hat a}, \mu_{a^*})$.

\subsection{Proof of Proposition~\ref{prop:linear}}\label{app:linear}

{\bf \underline{Proof of (\ref{eq:linear convergence})}:}  Suppose for a contradiction that (\ref{eq:linear convergence}) fails, i.e., 
\begin{equation}\label{eq:linear cont}
C^{{\rm lin}}_n(\mu, u,c, a^*)-C^{{\rm FB}}(u,c, a^*)=o(1/n).
\end{equation}
Thus, we can pick a sequence of IR and IC linear contracts $(w_n)$ with $w_n (x_n) = \frac{1}{n} \sum_{i=1} b_n (x_i)$ such that ${\mathbb E}_{a^*}[w_n (x^n)] -  C^{{\rm FB}}(u,c, a^*)=o(1/n)$. Note that
\begin{eqnarray*}
{\mathbb E}_{a^*}[w_n (x^n)] -  C^{{\rm FB}}(u,c, a^*)&=&{\mathbb E}_{a^*}[w_n (x^n)] -h({\mathbb E}_{a^*}[u(w_n (x^n))] )+h({\mathbb E}_{a^*}[u(w_n (x^n))] )  -h(c(a^*)) \\
&\geq& \frac{\inf_v h''(v)}{2} {\rm Var}_{a^*}[u(w_n (x^n))]+\underbrace{h({\mathbb E}_{a^*}[u(w_n (x^n))] )-h(c(a^*))}_{\geq 0 \text{ by IR}}
\end{eqnarray*}
by the variance-based estimate of the Jensen-inequality gap \citep{liao2018}. Thus, ${\mathbb E}_{a^*}[w_n (x^n)] -  C^{{\rm FB}}(u,c, a^*)=o(1/n)$ requires that ${\mathbb E}_{a^*}[u(w_n (x^n))]\to c(a^*)$ and ${\rm Var}_{a^*}[u(w_n (x^n))]=o(1/n)$ since $\inf_v h''(v) >0$. By a Taylor expansion argument, the latter implies ${\rm Var}_{a^*}[w_n (x^n)]=o(1/n)$.  As ${\rm Var}_{a^*}[w_n (x^n)]=\frac{1}{n}{\rm Var}_{a^*}[b_n(x_1)]$ for each $n$, this implies $\lim_{n\to\infty}{\rm Var}_{a^*}[b_n(x_1)]=0$. Thus, {${\mathbb E}_{a^*}[u(w_n (x^n))]\to c(a^*)$ implies that for any $\varepsilon>0$,  $\lim_{n\to\infty}\mu_{a^*}(\{x: |b_n(x)-h(c(a^*))|>\varepsilon\})=0$. This in turn implies that for all $a \neq a^*$, 
\begin{equation}\label{eq:lin-a}
\lim_{n\to\infty}\mu_{a}(\{x: |b_n(x)-h(c(a^*))|>\varepsilon\})= 0,
\end{equation}
because, by the data-processing inequality, $\mu_a(X')\ln\frac{\mu_a(X')}{\mu_{a^*}(X')}+\mu_a(X\setminus X')\ln\frac{\mu_a(X\setminus X')}{\mu_{a^*}(X\setminus X')} \leq {\rm KL}(\mu_{a}, \mu_{a^*})<\infty$ for all measurable $X'\subseteq X$.

The agent's expected payoff to choosing $a$ is then
\[
{\mathbb E}_a \left[u \left(\frac{1}{n}\sum_{i=1}^n b_n(x_i) \right) \right]-c(a) \geq {\mathbb E}_a \left[u(b_n(x_1)) \right]-c(a),
\]
by concavity of $u$. By (\ref{eq:lin-a}) and the fact that $u$ is bounded below, the $\limsup$ of the RHS as $n\to\infty$ is no less than $c(a^*)-c(a)$ . Thus, since ${\mathbb E}_{a^*}[u(w_n (x^n))]-c(a^*)\to 0$, (\ref{eq:IC'}) fails for any $a$ with $c(a)<c(a^*)$ at all large enough $n$.

{\bf \underline{``Moreover'' part}:} We show that (\ref{eq:linear convergence}) holds with equality if $\mu_{a^*}\not\in {\rm co}\{\mu_a: a\not=a^*\}$ and $\underline w$ is sufficiently low. We first exhibit a linear contract that achieves the first-best cost at $n=\infty$. 
This can be done by  finding $b: X\to [\underline w, \infty)$ such that (i) $\int b(x) d\mu_{a^*}(x)=h(c(a^*))$ and (ii) $\int b(x)d \mu_{a}(x)< h(c(a))$ for each $a\not= a^*$.
 Indeed, (i) ensures that (\ref{eq:IR}) binds, i.e., $u\left(\int b(x) d\mu_{a^*}(x)\right)-c(a^*)=0$. Given this, (ii) ensures that (\ref{eq:IC}) holds strictly, i.e., $u\left(\int b(x) d\mu_{a}(x)\right)- c(a)<0$ for each $a\not= a^*$. 
To find such a $b$, observe that, by a separation theorem \citep{dunford1988} and the assumption that $\mu_{a^*}\not\in{\rm co}\{\mu_a:a\not=a^*\}$, there is a bounded function $b:X\to \mathbb R$ with $\int b(x)d\mu_{a^*}(x)>\int b(x) d\mu_{a}(x)$ for all $a\not=a^*$. By scaling $b$ if necessary,  we can ensure $\int b(x) d\mu_{a^*}(x)-h(c(a^*))>\int b(x)d\mu_{a}(x)-h(c(a))$ for all $a\not=a^*$. By adding a constant  to $b$ if necessary, we can ensure $0=\int b(x) d\mu_{a^*}(x) -h(c(a^*))>\int b(x) \mu_{a}(x)-h(c(a))$ for all $a\not=a^*$. Since $b$ is bounded,  $b(X)\subseteq [\underline w, \infty)$ if $\underline w$ is sufficiently low.

To establish that (\ref{eq:linear convergence}) holds with equality, we modify the linear contract obtained above for each $n$ by setting $w_n (x^n)=\frac{1}{n} \sum_{i=1}^n (b(x_i) + \varepsilon_n) = \sum_x \nu_{x^n}(x)b(x)+\varepsilon_n$, where $\varepsilon_n\in\mathbb R_+$ is a constant and $\nu_{x^n}\in\Delta(X)$ denotes the realized empirical signal frequency. The constant $\varepsilon_n$ is pinned down for all large enough $n$ by requiring (\ref{eq:IR}) to bind, i.e., $\mathbb E_{a^*}[u(\sum_x \nu_{x^n}(x)b(x)+\varepsilon_n)]-c(a^*)=0$.

Note that $\lim_{n\to\infty}\varepsilon_n=0$ by the law of large numbers and construction of $b$. We moreover show that $\varepsilon_n=\frac{K}{n}+o(1/n)$ for some $K>0$. To see this, observe that
\begin{eqnarray*}
&& \inf_{w\in {\rm co} \{(b(x)+\varepsilon_n: x\in X\}} -\frac{u''(w)}{2}{\rm Var}_{a^*} \left[\sum_x \nu_{x^n}(x)b(x) \right]\\
 &\geq& u \left(\mathbb E_{a^*}\left[\sum_x \nu_{x^n}(x)b(x)+\varepsilon_n \right] \right)-\mathbb E_{a^*} \left[u \left(\sum_x \nu_{x^n}(x)b(x)+\varepsilon_n \right) \right] \\
 &\geq& u \left(\mathbb E_{a^*} \left[\sum_x \nu_{x^n}(x)b(x) \right] \right)+\varepsilon_n u' \left(\mathbb E_{a^*} \left[\sum_x \nu_{x^n}(x)b(x)+\varepsilon_n \right] \right) -c(a^*) \\
&=&  \varepsilon_n u' \left(\mathbb E_{a^*} \left[\sum_x \nu_{x^n}(x)b(x)+\varepsilon_n \right] \right),
\end{eqnarray*}
where the first inequality uses the variance-based estimate of the Jensen-inequality gap (note $\inf -u'' > 0$ on the relevant domain of $w$, as $b$ is bounded), the second inequality uses the concavity of $u$ and construction of $\varepsilon_n$, and the last equality uses the construction of $b$. Since the variance term in the first line is $o(\frac{1}{n})$, we have $\varepsilon_n=\frac{K}{n}+o(\frac{1}{n})$ for some $K>0$. 

Because $\varepsilon_n\to 0$ and the original linear contract satisfies (\ref{eq:IC}) strictly at $n=\infty$, the modified contracts satisfy (\ref{eq:IC}) for all large enough $n$. Under these contracts, the principal's cost difference relative to the first-best cost is $\varepsilon_n = \frac{K}{n}+o(\frac{1}{n})$, as required.

\subsection{Proof of Proposition~\ref{prop:limit2}}\label{sec:app-limit}

We prove the following generalization of Proposition~\ref{prop:limit2} to an arbitrary finite action space. As usual, we write contracts $v_n$ as functions of log-likelihood scores.

\begin{prop}\label{prop:limit3}
For any weakly convergent sequence of optimal contracts $(v^*_n)$, there exists $\gamma: A^- \to\mathbb R$ such that
\[
\lim_{n\to\infty} v^*_{n} ( L)=
\begin{cases}
c(a^*) &\text{ if }  L(a)>\gamma(a) \text{ for each } a \in A \setminus \{a^*\}  \\
u(\underline w) &\text{ if }  L(a)< \gamma(a) \text{ for some } a \in A \setminus \{a^*\},
\end{cases}
\]
where $\gamma ( \hat a) = \mathbb{E}_{\hat a} [L_n (\hat a)]$ is maximally lenient for all $\hat a\in\argmin_{a \in A^{-}}{\rm KL}(\mu_{a}, \mu_{a^*})$. 
\end{prop}

\subsubsection{The Limit Contract is Binary} 
  
By Appendix~\ref{app:optimal}, for all large enough $n$,  $v^*_n$ is given by (\ref{eq:v*}) for multipliers $\lambda_n$ and $\kappa_n(a)$ for each $a\not=a^*$. 
Since $(v^*_n)$ approximates the first-best, $\lim_{n\to\infty}{\mathbb E}_{a^*}[v^*_n( L_n)]=c(a^*)$. Since $L_n$ converges in probability to ${\mathbb E}_{a^*}[ L_n] > 0$, (\ref{eq:v*}) implies 
\begin{equation}\label{eq:limit multipliers}
\lim_{n\to\infty} \left(\lambda_n+\sum_{a\not=a^*}\kappa_n(a) \right)=h'(c(a^*)).
\end{equation}   
Up to restricting to an appropriate subsequence, we can assume that $\lim_{n\to\infty}\frac{1}{n}\ln \kappa_{n}(a)=:\gamma(a)$ exists (allowing for $-\infty$) for each $a\not=a^*$. Since $\kappa_n(a)$ is bounded by (\ref{eq:limit multipliers}),
\[
\lim_{n\to\infty}\kappa_n(a) \left(1-\exp[-n L(a)]\right)=
\begin{cases}
\lim_{n\to\infty}\kappa_n(a) &\text{ if }  L(a)>\gamma(a) \\
-\infty &\text{ if }  L(a)<\gamma(a).
\end{cases}
\]
Thus, by (\ref{eq:v*}), 
\begin{equation}\label{eq:bin-lim}
\lim_{n\to\infty}v^*_{n}( L) 
=  \begin{cases}
c(a^*) &\text{ if }  L(a)>\gamma(a) \text{ for each } a \in A \setminus \{a^*\}  \\
u(\underline w) &\text{ if }  L(a)< \gamma(a) \text{ for some } a \in A \setminus \{a^*\} .
\end{cases}
\end{equation}

\subsubsection{Maximal Leniency of Limit Thresholds}
 
It remains to show that in (\ref{eq:bin-lim}), $\gamma(\hat a)=\mathbb{E}_{\hat a} [ L_n (\hat a)]$ for all $\hat a\in\argmin_{a\in A^-}{\rm KL}(\mu_a, \mu_{a^*})$. 

By the weak convergence of $(v^*_n)$ and (\ref{eq:bin-lim}), for any $\varepsilon>0$ there is $n_\varepsilon$ such that $v^*_n(L)\leq u(\underline w)+\varepsilon$ if $L(a)\leq \gamma(a)-\varepsilon$ for some $a\not=a^*$ and $n\geq n_\varepsilon$. Thus
 \begin{eqnarray*}
 &&\liminf_n \frac{1}{n}\ln {\rm Var}_{a^*}[v^*_n( L_n)] =  \liminf_n \frac{1}{n}\ln {\mathbb E}_{a^*}[({\mathbb E}_{a^*}[v^*_n(L_n)]-v^*_n( L_n))^2]  \\
&\geq&  \liminf_n \frac{1}{n}\ln {\mathbb P}_{a^*}[ L_n(a)<\gamma(a)-\varepsilon \text{ for some } a\not=a^*] (c(a^*)-u(\underline w)-2\varepsilon)^2 \\
  &=&\liminf_n \frac{1}{n}\ln {\mathbb P}_{a^*}[ L_n(a)<\gamma(a)-\varepsilon \text{ for some } a\not=a^*] 
\geq  -  \min_{a \neq a^*} \inf_{\ell<\gamma(a)-\varepsilon}I_{a^*, a}(\ell)
\end{eqnarray*}
where the first inequality uses that ${\mathbb E}_{a^*}[v^*_n(L_n)]\geq c(a^*)-\varepsilon$ for all large enough $n$ and the second inequality uses Cram\'er's theorem (\ref{eq:cramer}).  Since $\varepsilon>0$ is arbitrary, we have $ \liminf_n \frac{1}{n}\ln {\rm Var}_{a^*}[v^*_n( L_n)] \geq -  \min_{a \neq a^*} \inf_{\ell<\gamma(a)}I_{a^*, a}(\ell)$.

Since ${\rm Var}_{a^*}[v^*_n( L_n)]$ vanishes at exponential rate $\min_{a\in A^{-}}{\rm KL}(\mu_{a}, \mu_{a^*})$ (by the proof of Theorem~\ref{thm:main}), it follows that (using (\ref{eq:variational})) that
\begin{equation}\label{eq:HPs}
\min_{a\in A^{-}}{\rm KL}(\mu_{a}, \mu_{a^*})\leq\min_{a \neq a^*} \inf_{\ell<\gamma(a)}I_{a^*,a}(\ell)= \inf_{\substack{a\not=a^*, \, \nu\in\Delta(X) \text{ s.t. } \\ \int \ln\frac{d\mu_{a^*}}{d\mu_{a}}(x)d\nu(x) <\gamma (a)}}{\rm KL}(\nu, \mu_{a^*}).
\end{equation}

Suppose for a contradiction that $\gamma(\hat a) \neq \mathbb{E}_{\hat a} [ L_n (\hat a)]$ for some $\hat a \in \argmin_{a\in A^-}{\rm KL}(\mu_a, \mu_{a^*})$. If $\gamma(\hat a) > \mathbb{E}_{\hat a} [ L_n (\hat a)]$, then since $\mathbb{E}_{\hat a} [ L_n (\hat a)]=-{\rm KL}(\mu_{\hat a}, \mu_{a^*})$ by definition of the log-likelihood score, Lemma~\ref{lem:KL} and the strict convexity of KL divergence implies that 
\[
 {\rm KL}(\mu_{\hat a}, \mu_{a^*})>\inf_{\nu\in\Delta(X) \text{ s.t. } \int \ln\frac{\mu_{a^*}(x)}{\mu_{\hat a}(x)}d\nu(x) <\gamma(\hat a)}{\rm KL}(\nu, \mu_{a^*}),
 \] 
which contradicts (\ref{eq:HPs}). If instead $\gamma(\hat a) <\mathbb{E}_{\hat a} [ L_n (\hat a)]$, there exists $a\not=a^*$ such that
\[
\gamma(a) \geq \mathbb{E}_{\hat a} [ L_n (a)]=\int \ln\frac{d\mu_{a^*}}{d\mu_{a}}(x)d\mu_{\hat a}(x),
\]
as otherwise $\lim_{n\to\infty}{\mathbb P}_{\hat a}[ L_n(a)>\gamma(a) \text{ for all } a\not=a^*]=1$ by the weak law of large numbers, which leads to a violation of (\ref{eq:IC'}) at all large enough $n$. Let $B:=\{\nu\in\Delta(X): {\rm KL}(\nu, \mu_{a^*})\leq {\rm KL}(\mu_{\hat a}, \mu_{a^*})\}$. Then (\ref{eq:HPs}) implies that, for every $\nu\in B$,
$\gamma(a) \leq \int\ln\frac{d\mu_{a^*}}{d\mu_{a}}(x)d\nu(x)$.\footnote{If $\gamma(a) > \int \ln\frac{d\mu_{a^*}}{d\mu_{a}}(x)d\nu(x)$, one can find $\nu'$ close to $\nu$ with ${\rm KL}(\nu', \mu_{a^*})<{\rm KL}(\mu_{\hat a}, \mu_{a^*})$ and $\gamma(a) > \int \ln\frac{d\mu_{a^*}}{d\mu_{a}}(x)d\nu'(x)$ (based on the convexity of KL divergence), which violates (\ref{eq:HPs}). }  Thus, $\gamma(a) =\int\ln\frac{d\mu_{a^*}}{d\mu_{a}}(x)d\mu_{\hat a}(x)$, as $\mu_{\hat a}\in B$. By the strict convexity of KL divergence, this shows that $\mu_{\hat a}$ is the unique point at which $B$ is tangent to the hyperplane $\{\nu\in\Delta (X): \gamma(a) = \int \ln\frac{d\mu_{a^*}}{d\mu_{a}}(x) d\nu(x)\}$. Likewise, by construction of $B$ and Lemma~\ref{lem:KL}, $\mu_{\hat a}$ is the unique element in $B$ that is tangent to the hyperplane $\{\nu\in\Delta (X): {\rm KL}(\mu_{\hat a}, \mu_{a^*}) = \int \ln\frac{\mu_{a^*}(x)}{\mu_{\hat a}(x)}d\nu(x) \}$.  By $\mu_{a^*}\not=\mu_{a}$, these two hyperplanes are not parallel to each other. As every boundary point of $B$ has a unique tangent space (by strict convexity of KL divergence), this is a contradiction.

\subsection{Proof of Theorem~\ref{thm:severe}}\label{app:severe}

We prove the following generalization of Theorem~\ref{thm:severe} that allows for non-binary actions. 
For simplicity, we assume there is a unique action $\hat a$ that minimizes $c$. We consider the case where Assumption~\ref{asp}.2 is violated, in the sense that $c(\hat a) \leq u(\underline w)$. Then the first-best cost is
$C^{\rm{FB}}(u,c, a^*)= h \left(u(\underline w)+c(a^*)-c(\hat a) \right)$. As in the main text,  we assume that $u(\underline w)+c(a^*)-c(\hat a)<\lim_{w\to\infty}u(w)$, so $C^{\rm{FB}}(u,c, a^*)$ is well-defined.

\begin{thm} Under both general and binary contracts, the second-best cost converges to the first-best exponentially at rate $\rho(\mu):=\min\{{\rm Ch}(\mu_{\hat a}, \mu_{a^*}), \min_{a\in A^-\setminus\{\hat a\}}{\rm KL}(\mu_a, \mu_{a^*}) \}$:
\[
C^{\rm{SB}}_n(\mu, u,c, a^*)-C^{\rm{FB}}(u,c, a^*)=\exp[-\rho(\mu) n+o(n)];
\]
\[
C^{\rm{bin}}_n(\mu,  u,c, a^*)-C^{\rm{FB}}(u,c, a^*)=\exp[-\rho(\mu)n+o(n)].
\]
\end{thm}

To understand the convergence rate $\rho(\mu)$, it is helpful to distinguish two cases.  First, suppose ${\rm Ch}(\mu_{\hat a}, \mu_{a^*})<\min_{a\in A^-\setminus\{\hat a\} }{\rm KL}(\mu_a, \mu_{a^*})$; intuitively, a deviation to the least costly action $\hat a$ is relatively harder to detect than other deviations. In this case, we show based on a similar intuition as in the binary-action case (Theorem~\ref{thm:severe}) that the convergence rate is ${\rm Ch}(\mu_{\hat a}, \mu_{a^*})$.  
Second, suppose ${\rm Ch}(\mu_{\hat a}, \mu_{a^*})>\min_{a\in A^-\setminus\{\hat a\}}{\rm KL}(\mu_a, \mu_{a^*})$; intuitively, a deviation to $\hat a$ is easier to detect than some other deviations.
 In this case, we show based on a similar logic as in Theorem~\ref{thm:main} that the convergence rate is given by $\min_{a\in A^-\setminus\{\hat a\} }{\rm KL}(\mu_a, \mu_{a^*})$.

\subsubsection{Upper Bound on Inefficiency under Binary Contracts}

 We first show that, as $n\to\infty$,  $C^{\rm bin}_n(\mu, u, c, a^*)-C^{\text{FB}}(u, c, a^*)$ vanishes at least as fast as at the exponential rate $\rho (\mu)$. For each $a \in A^-$, fix some $\gamma(a) \in ({\mathbb E}_a[ L_n(a)],{\mathbb E}_{a^*}[ L_n(a)])$. For each $a'\in \{a^*\}\cup A^-$, define $p_{a', n}:={\mathbb P}_{a'}[ L_n(a)\geq \gamma(a) \forall a\in A^-]$. Observe that, by the weak law of large numbers and the choice of $\gamma(a)$, we have
\[
\lim_{n\to\infty}p_{a^*, n}=1 \quad  \text{ and } \quad \lim_{n\to\infty}p_{a, n}=0  \text{ for all } a \in A^- .
\]
Consider the sequence of binary contracts $(v_n)$ given by
\[
v_n (x^n)= \begin{cases}
v^+_n &\text{ if }   L_n(a)\geq \gamma(a)   \text{ for all  } a\in A^- \\
u(\underline w)  &\text{ otherwise,} 
\end{cases}
\]
where $v_{n}^+ :=u(\underline w)+\frac{c(a^*)-c(\hat a)}{p_{a^*,n}-p_{\hat a,n}}$. 

Note that, up to restricting to large enough $n$, these contracts are well-defined, because $\lim_{n\to\infty}v_n^+=u(\underline w)+c(a^*)-c(\hat a)$. Moreover, by choice of $v_{n}^+$, we have ${\mathbb E}_{a^*}[v_n(x^n)]-c(a^*)= {\mathbb E}_{\hat a}[v_n (x^n)]-c(\hat a)$, i.e., (\ref{eq:IC'}) holds with equality at deviation $\hat a$. 
Furthermore, $\lim_{n\to\infty}{\mathbb E}_{a^*}[v_n(x^n)]-c(a^*)=u(\underline w)-c(\hat a) > \lim_{n\to\infty}{\mathbb E}_{a}[v_n(x^n)]-c(a)=u(\underline w)-c(a)$ for all $a\in A^- \setminus \{\hat a\}$, so deviations to any other action in $A^-$ are also suboptimal at large enough $n$. Finally, the agent's expected payoff to choosing any $a \in A^+$ is bounded above by $u(\underline w)+c(a^*)-c(\hat a)-c(a)<u(\underline w)-c(\hat a)=\lim_{n\to\infty}{\mathbb E}_{a^*}[v_n(x^n)]-c(a^*)$.  Thus, (\ref{eq:IC'}) holds for all large enough $n$.

For all large enough $n$, the difference between the principal's cost under $v_n$ and the first-best can be decomposed as
\begin{eqnarray*}
&&{\mathbb E}_{a^*}[h(v_n (x^n))]-h(u(\underline w)+c(a^*)-c(\hat a))\\
&=&{\mathbb E}_{a^*}[h(v_n (x^n))]-h({\mathbb E}_{a^*}[v_n (x^n)])+h({\mathbb E}_{a^*}[v_n (x^n)])-h(u(\underline w)+c(a^*)-c(a)) \\
&= &{\mathbb E}_{a^*}[h(v_n (x^n))]-h({\mathbb E}_{a^*}[v_n (x^n)])+h\left({\mathbb E}_{\hat a}[v_n (x^n)]+c(a^*)-c(\hat a)\right)-h(u(\underline w)+c(a^*)-c(\hat a)),
\end{eqnarray*}
where the second equality uses the binding (\ref{eq:IC'}) against $\hat a$.

As in the proof of Theorem~\ref{thm:main}, ${\mathbb E}_{a^*}[h(v_n (x^n))]-h({\mathbb E}_{a^*}[v_n(x^n)])$ vanishes at least as fast as at the exponential rate $
\min_{a\in A^-} \inf_{\ell\leq \gamma(a)}I_{a^*,a}(\ell)$. Moreover, for each $a \in A^- $, Lemma~\ref{lem:KL}, (\ref{eq:variational}), and the continuity of $I_{a^*, a}$ implies that
\[
\sup_{\gamma> {\mathbb E}_{a}[ L_n(a)]} \inf_{\ell\leq\gamma}I_{a^*,a}( \ell)=\inf_{\ell\leq {\mathbb E}_{a}[ L_n(a)]}I_{a^*,a}( \ell) ={\rm KL}(\mu_a, \mu_{a^*})  \quad \text{ and } \quad \inf_{\ell\leq 0}I_{a^*,a}(\ell) ={\rm Ch}(\mu_a, \mu_{a^*}).
\]
Thus, by choosing $\gamma(a)$ to be arbitrarily close to the maximally lenient threshold ${\mathbb E}_{a}[ L_n(a)]$ for each $a \in A^- \setminus \{ \hat a\}$ and setting $\gamma(\hat a)=0$, ${\mathbb E}_{a^*}[h(v_n(x^n))]-h({\mathbb E}_{a^*}[v_n(x^n)])$ can be made to vanish at a rate arbitrarily close to the exponential rate $\rho(\mu)$.

Next, note that the rate at which $h({\mathbb E}_{\hat a}[v_n(x^n)]+c(a^*)-c(\hat a))-h(u(\underline w)+c(a^*)-c(\hat a))$ vanishes is the same as the rate at which ${\mathbb P}_{\hat a}[ L_n(a)\geq\gamma(a) \, , \forall a\in A^-]$ vanishes. Moreover, ${\mathbb P}_{\hat a}[ L_n(a)\geq\gamma(a) \, ,\forall a\in A^-]\leq {\mathbb P}_{\hat a}[ L_n(\hat a)\geq\gamma(\hat a)]$, and the latter vanishes at rate ${\rm Ch}(\mu_{\hat a}, \mu_{a^*})$ since we set $\gamma(\hat a)=0$.

The previous two paragraphs imply the desired conclusion.

\subsubsection{Lower Bound on Inefficiency under General Contracts}

Next, we show that $C_n(\mu, u, c, a^*)-C^{\text{FB}}(u, c, a^*)$ vanishes at most as fast at the exponential rate $\rho (\mu)$. We consider two cases:

\underline{\bf Case (i): $\rho(\mu)=\min_{a\in A^-\setminus\{\hat a\}}{\rm KL}(\mu_a, \mu_{a^*})$}. Take any $a'\in \argmin_{a\in A^-\setminus\{\hat a\}}{\rm KL}(\mu_a, \mu_{a^*})$. Consider the relaxed problem where the agent's action space is $A=\{a^*, a'\}$ and his outside option yields payoff $u(\underline w)-c(\hat a)$. This is indeed a relaxation of the original problem, as the agent's expected utility of choosing $\hat a$ is no less than $u(\underline w)-c(\hat a)$. Take a sequence of optimal contracts $(v_n)$ under the relaxed problem, where we write each $v_n$ as a weakly increasing function of $L_n :=L_n(a')$. Suppose for a contradiction that the principal's cost in this problem converges to the first-best $h(u(\underline w)+c(a^*)-c(\hat a))$ faster than at some exponential rate $\lambda>\rho(\mu) = {\rm KL}(\mu_{a'}, \mu_{a^*})$.  Observe that 
\begin{eqnarray*}
&&{\mathbb E}_{a^*}[h(v_n(L_n))]-h(u(\underline w)+c(a^*)-c(\hat a))\\
&=&{\mathbb E}_{a^*}[h(v_n(L_n))]-h({\mathbb E}_{a^*}[v_n(L_n)])+h({\mathbb E}_{a^*}[v_n(L_n)])-h(u(\underline w)+c(a^*)-c(\hat a)) \\
&\geq & {\mathbb E}_{a^*}[h(v_n(L_n))]-h({\mathbb E}_{a^*}[v_n(L_n)]),
\end{eqnarray*}
where the inequality uses the (\ref{eq:IR'}) constraint of the relaxed problem. Thus, ${\mathbb E}_{a^*}[h(v_n(L_n))]-h({\mathbb E}_{a^*}[v_n(L_n)])$ vanishes faster than at rate $\lambda>{\rm KL}(\mu_{a'}, \mu_{a^*})$. As in Appendix~\ref{app:lowerbd}, this implies that (\ref{eq:IC'}) fails for all large enough $n$, a contradiction.

\underline{\bf Case (ii): $\rho(\mu)= {\rm Ch}(\mu_{\hat a}, \mu_{a^*})$}. Consider the relaxed problem with action space $A=\{a^*, \hat a\}$ and outside option $0$. Take a sequence of optimal contracts $(v_n)$ under the relaxed problem, where we write each $v_n$ as a weakly increasing function of $L_n :=L_n(\hat a)$.  Let $\hat L_a:={\mathbb E}_a[L_n]$ for $a \in \{a^*,\hat a\}$. 
Suppose for a contradiction that the principal's cost in this problem converges to the first-best $h(u(\underline w)+c(a^*)-c(\hat a))$ faster than at some exponential rate $\lambda>\rho(\mu) = {\rm Ch}(\mu_{\hat a}, \mu_{a^*})$.  Observe that 
\begin{eqnarray*}
&&{\mathbb E}_{a^*}[h(v_n(L_n))]-h(u(\underline w)+c(a^*)-c(\hat a))\\
&=&{\mathbb E}_{a^*}[h(v_n(L_n))]-h({\mathbb E}_{a^*}[v_n(L_n)])+h({\mathbb E}_{a^*}[v_n(L_n)])-h(u(\underline w)+c(a^*)-c(\hat a)) \\
&\geq & {\mathbb E}_{a^*}[h(v_n(L_n))]-h({\mathbb E}_{a^*}[v_n(L_n)])+h({\mathbb E}_{\hat a}[v_n(L_n)]+c(a^*)-c(\hat a))-h(u(\underline w)+c(a^*)-c(\hat a)),
\end{eqnarray*}
where the inequality uses (\ref{eq:IC'}). Thus, both ${\mathbb E}_{a^*}[h(v_n(L_n))]-h({\mathbb E}_{a^*}[v_n(L_n)])$ and $h({\mathbb E}_{\hat a}[v_n(L_n)]+c(a^*)-c(\hat a))-h(u(\underline w)+c(a^*)-c(\hat a))$ vanish faster than at rate $\lambda>{\rm Ch}(\mu_{\hat a}, \mu_{a^*})$. The latter also implies that ${\mathbb E}_{\hat a}[v_n(L_n)]-u(\underline w)$ vanishes faster than at rate $\lambda$.

By (\ref{eq:variational}) and (\ref{eq:Ch-lem}), we have $I_{a^*, \hat a}(0)=I_{\hat a, \hat a}(0)={\rm Ch}(\mu_{\hat a}, \mu_{a^*})$. Thus, by the continuity of the rate functions, there is an interval $\Gamma \ni 0$ such that $\lambda>I_{a^*, \hat a}(\ell), I_{\hat a, \hat a}(\ell)$ for all $\ell\in \Gamma$. By the same argument as in Appendix~\ref{app:lowerbd}, in order for ${\mathbb E}_{a^*}[h(v_n(L_n))]-h({\mathbb E}_{a^*}[v_n(L_n)])$ to vanish faster than at rate $\lambda$, we must have $\lim_{n\to\infty} v_n(\hat L_{a^*})-v_n(0)=0$. Thus, $u(\underline w)+c(a^*)-c(\hat a)=\lim_{n\to\infty}v_n(\hat L_{a^*})=\lim_{n\to\infty}v_n(0)$. By an analogous argument, in order for ${\mathbb E}_{\hat a}[v_n(L_n)]-u(\underline w)$ to vanish faster than at rate $\lambda$, we must have $\lim_{n\to\infty} v_n(0)-u(\underline w)=0$. This contradicts $c(a^*)>c(\hat a)$.

\subsection{Details for Section~\ref{sec:general}}\label{app:general}

\subsubsection{Non-i.i.d.\ Signals}\label{app:non-iid}

Consider a sequence $\mu = (\mu^n)$ of monitoring technologies indexed by $n$.  Fix a measurable signal space $Z$ that is common across $n$ (without loss of generality) and endowed with some $\sigma$-finite measure $\nu$. For each $n$ and $a \in A$, the monitoring technology $\mu^n$ specifies a (not necessarily full-support) signal distribution $\mu^n_a \in \Delta (Z)$ that is absolutely continuous with respect to $\nu$, so that $\mu^n_a$ admits a density function $g^n_a$.  Assume that for each $a,a'\in A \setminus \{a^*\}$, the log-likelihood score $L_n(a'):=\frac{1}{n}\ln\frac{g^n_{a^*}(z)}{g^n_{a'}(z)}$ is well-defined $\mu^n_{a}$-almost surely.

The remaining setup and assumptions on payoffs are the same as in Section~\ref{sec:model}. In particular, the principal's second-best cost solves
\[
C^{\rm{SB}}_n(\mu, a^*, u, c)=\inf_{w:Z\to [\underline w, \infty)} \int w(z) \, d\mu_{a^*}^n(z)
\]
subject to the IC and IR constraints
\begin{equation}\tag{IC}\label{eq:IC''}
 \int u(w(z))  \, d\mu_{a^*}^n(z)-c(a^*) \geq  \int u(w(z)) \, d\mu_{a}^n(z)-c(a), \, \forall a \in A,
\end{equation}
\begin{equation}\tag{IR}\label{eq:IR''}
 \int u(w(z))  \, d\mu_{a^*}^n(z)-c(a^*)\geq 0.
\end{equation}

The following key requirement captures that monitoring technologies become very precise as $n \to \infty$: We assume that, under any chosen action $a\in A$, the distributions of the log-likelihood score $L_n(a')$ for each $a'\not=a^*$ obey the \textbf{\textit{large-deviation principle}} with respect to some \textbf{\textit{rate function}} $I_{a, a'}:\mathbb R\to \overline{\mathbb R}_+$ as $n \to \infty$. That is, for each %$a\in A$, $a'\not=a^*$ and 
 measurable set $B\subseteq\mathbb R$,  
\begin{equation}\label{eq:LDP}
\begin{split}
-\inf_{{\ell}\in{\rm int} B}I_{a, a'}(\ell)\leq &\liminf_{n\to\infty}\frac{1}{n}\ln {\mathbb P}_{a}[ L_n(a')\in B]\\
\leq &\limsup_{n\to\infty}\frac{1}{n}\ln {\mathbb P}_a[ L_n(a')\in B]\leq -\inf_{{\ell}\in{\rm cl} B}I_{a, a'}({\ell}).
\end{split}
\end{equation}
Intuitively, $I_{a, a'}(\ell)$ measures how atypical it is, under the chosen action $a$, for the score $L_n (a')$ to take value $\ell$ when $n$ is large. We impose an identification condition: For each $a\in A$ and $a'\not=a^*$, $I_{a, a'}$ is uniquely minimized by some value $\hat L_a(a')$, where $\hat L_a(a')\not=\hat L_{\hat a}(a')$ for $a\not=\hat a$. Thus, the distribution of $ L_n(a')$ under $a$ concentrates on the deterministic limit $\hat L_a(a')$ as $n$ becomes large, where the limit is distinct across different actions $a$, thus allowing the principal to identify the agent's chosen action as $n \to \infty$. We also impose the following regularity conditions: For each $a\in A$ and $a'\not=a^*$, $I_{a,a'}$ has compact level sets,\footnote{That is, $\{ L\in\mathbb R:  I_{a, a'}( L)\leq k\}$ is compact for each  $k\in\mathbb R_+$.} and $I_{a^*, a'}$ is continuous at $\hat L_{a'}(a')$ and $\hat L_{a^*}(a')$.

This setting encompasses several natural examples of rich/precise monitoring data:

\begin{ex}[Many signals]
Suppose $Z=\bigcup_n X^n$, i.e., $n$ represents the number of signals observed by the principal, as in our main model. However, beyond i.i.d.\ signals, we can allow for correlated signals: The signal sequence $(x_1,\ldots, x_n)$ is generated by a Markov transition matrix $M_a(\cdot|x)\in\Delta(X)$ that depends on the chosen action $a$. Assume for simplicity that $X$ is finite, each $M_a(\cdot | x)$ has full support, and the initial signal $x_1$ is drawn from some full-support distribution that can depend on $a$ and $n$.    

This setting satisfies our assumptions, with a rate function that is the solution to the following optimization problem \cite[e.g.,][]{dembo2009}:
\[
I_{a, a'}(\ell)=\inf_{\nu\in\Delta(X\times X) }\sum_x\nu(x){\rm KL} \left(\nu(\cdot|x), M_a(\cdot|x)\right) \quad \text{ subject to }
\]
\[
\sum_{x,x'}\nu(x,x')\ln\frac{M_{a^*}(x'|x)}{M_{a'}(x'|x)} = \ell \quad \text{ and } \quad \sum_{x'\in X}\nu(x,x')=\sum_{x'\in X}\nu(x',x) \text{ for each } x\in X,
\]
where $\nu(x):=\sum_{x''\in X}\nu(x,x'')$ for each $x$ and $\nu(x'|x):=\frac{\nu(x,x')}{\nu(x)}$ for each $x,x'\in X$ with $\nu(x)>0$. In our main model where each $M_a(\cdot|x)$ is independent of $x$, this expression reduces to the rate function (\ref{eq:variational}) from Cram\'er's theorem.\footnote{This is because in this case it is without loss of generality to minimize over product measures $\nu\in\Delta(X\times X)$, due to the convexity of KL divergence.} \finex
\end{ex}

We can also capture a principal who, instead of observing many signals, observes a single signal, which becomes increasingly precise as $n \to \infty$:

\begin{ex}[Vanishing observation noise]\label{ex:gaussian}
Suppose $A\subseteq \mathbb R$. If the agent chooses action $a$, the principal observes signal $x=a+\kappa_n\varepsilon$, where $\varepsilon$ is a random shock that follows the standard normal distribution.\footnote{We conjecture that we can also allow for shocks $\varepsilon$ that follow other distributions (subject to some assumptions).} Under the scaling factor $\kappa_n=\frac{1}{\sqrt{n}}$, we have
\[
I_{a, a'}(\ell)= \inf_{x\in\mathbb R} \frac{(x-a)^2}{2} \quad \text{ subject to } \quad \frac{(x-a')^2-(x-a^*)^2}{2}=\ell. \finex
\]
\end{ex}

The following result extends Theorem~\ref{thm:main} to the current setting:

\begin{thm}\label{thm:general} Under both general and binary contracts, the second-best cost converges to the first-best exponentially at rate $\rho(\mu):=\min_{a \in A^-}I_{a^*, a}(\hat L_a(a))$:
\[
C^{\rm{SB}}_n(\mu, a^*, u,c)-C^{{\rm FB}}_n(a^*, u,c)=\exp[-\rho(\mu) n+o(n)];
\]
\[
C^{{\rm bin}}_n(\mu, a^*, u,c)-C^{{\rm FB}}_n(a^*, u,c)=\exp[- \rho(\mu)n+o(n)].
\]
\end{thm}

To understand the convergence rate, recall that, for each $a\not=a^*$, $I_{a^*, a}(\hat L_a(a))$ measures how atypical it is, if the chosen action is $a^*$, to observe the score value $L_n(a)=\hat{L}_a(a)$ (i.e., the limit score under a deviation to $a$). Thus, $\rho (\mu) :=\min_{a \in A^-}I_{a^*, a}(\hat L_a(a))$ is again a measure of the detectability of the hardest-to-detect deviation. Under i.i.d.\ signals, $I_{a^*, a}(\hat L_a(a))={\rm KL}(\mu_a, \mu_{a^*})$, so Theorem~\ref{thm:general} reduces to Theorem~\ref{thm:main}. The proof of Theorem~\ref{thm:general} follows similar steps as that of Theorem~\ref{thm:main} and is presented in Online Appendix~\ref{app:noniid-pf}. To achieve the optimal convergence rate, we again construct binary contracts based on the maximally lenient thresholds $\gamma(a)\searrow \hat L_a(a)$ for each $a\in A^{-}$.

\subsubsection{Adjustable Actions}\label{app:adjust}

 Fix $T \in \mathbb{N} \setminus \{0\}$ and assume that the agent sequentially chooses $T$ actions, $a_1, \ldots, a_T \in A$. Each action $a_t$ generates $\frac{n}{T}$ i.i.d.\ draws of signals, $x^n_t=(x_{t,1},\ldots, x_{t, \frac{n}{T}}) \in X^{\frac{n}{T}}$, from $\mu_{a_t} \in \Delta(X)$, where we restrict attention to $n$ that are divisible by $T$ and impose the same assumptions on the monitoring technology $\mu$ as in Section~\ref{sec:model}. Signals are observed by both the principal and agent. Thus, a strategy of the agent is described by a collection of mappings $\alpha=(\alpha_t)_{t=1}^T$, where $\alpha_t: X^{\frac{(t-1)n}{T}}\to A$ specifies the agent's action $a_t$ after each history of past signal realizations $(x^n_1,\ldots, x_{t-1}^n)$ (and $x_0^n$ denotes the empty history).

We continue to assume that the agent has additively separable payoffs, with a consumption utility $u: [\underline w, \infty) \to \mathbb{R}$ over money and a total cost $\frac{1}{T} \sum_{t=1}^T c(a_t)$ over action sequences $a_1, \ldots, a_T$, where $u$ and $c$ satisfy Assumption~\ref{asp}.1--2. The principal designs a contract $v: X^{n}\to [\underline w, \infty)$ that specifies a one-shot utility payment $v(x^n)$ as a function of the entire sequence $x^n = (x^n_1, \ldots, x^n_T)$ of $n$ signals. In the current setting, the principal may want the agent to choose different actions $a_t$ depending on past signal realizations  $(x^n_1,\ldots, x^n_{t-1})$. To accommodate this, instead of exogenously fixing a target action profile, we assume that the risk-neutral principal receives a monetary payoff $\frac{1}{T} \sum_{t=1}^T g(a_t)$ from the agent's actions, for some function $g:A\to\mathbb R$.

In the second-best problem, the principal jointly chooses a contract $v$ and agent-strategy $\alpha$ to maximize her payoff,
\[
G^{\rm{SB}}_n  :=\sup_{v, \alpha}  \, {\mathbb E}_{\alpha}\left[\frac{1}{T} \sum_{t=1}^T g(a_t) - h(v(x^n))\right],
\]
subject to the IC and IR constraints
\begin{equation}\tag{IC}\label{eq:IC-adj}
{\mathbb E}_{\alpha}\left[v(x^n)- \frac{1}{T}\sum_{t=1}^T c(a_t) \right] \geq {\mathbb E}_{\alpha'}\left[v(x^n)- \frac{1}{T}\sum_{t=1}^T c(a_t) \right], \; \forall \alpha',
\end{equation}
\begin{equation}\tag{IR}\label{eq:IR-adj}
{\mathbb E}_{\alpha}\left[v(x^n)- \frac{1}{T}\sum_{t=1}^T c(a_t) \right] \geq 0.
\end{equation}
Let $G^{\rm bin}_n $ denote the principal's value under the restriction to binary contracts.

As before, we also consider the first-best problem, which corresponds to only imposing (\ref{eq:IR-adj}). We assume there is $a^*\in A$ such that it is strictly optimal to set $\alpha_t(\cdot)=a^*$ for all $t$ under the first-best problem at each $n$.\footnote{For some specifications of $g$, $u$, and $c$, the first-best solution may require the agent to choose different actions $a_t$ depending on past signal realizations. We rule out these cases to keep the departure from the main model parsimonious.} Thus, the principal's first-best payoff is $G^{\rm FB}=g(a^*)-h(c(a^*))$.

Given $a^*$, we impose Assumption~\ref{asp}.3 and let $A^-:=\{a\in A: c(a)<c(a^*)\}$ and $A^+:=\{a\in A: c(a)>c(a^*)\}$. For technical convenience, we also impose the following richness condition on $\mu$: For each $p\in (0,1)$ and $a\in A$, there is a set of signals $Y\subseteq X$ such that $\mu_{a}(Y)=p$.\footnote{This is satisfied, for example, if we augment the monitoring technology $\mu$ by an uninformative signal that is drawn uniformly (independently of the agent's action).}

The following result fixes $T$ and characterizes the optimal convergence rate to the first-best in the limit as $n \to \infty$:
\begin{thm}\label{prop:adjusment} Under both general and binary contracts, the second-best payoff converges to the first-best exponentially at rate $\frac{1}{T}\min_{a \in A^-}{\rm KL} (\mu_a, \mu_{a^*})$ as $n\to\infty$:
\[
G^{\rm FB}-G^{\rm{SB}}_n =\exp[- \frac{1}{T}\min_{a\in A^-}{\rm KL}(\mu_a, \mu_{a^*})n+o(n)];
\]
\[
G^{\rm FB}-G^{\rm bin}_n =\exp[- \frac{1}{T}\min_{a\in A^-}{\rm KL}(\mu_a, \mu_{a^*})n+o(n)].
\]
\end{thm}
Theorem~\ref{prop:adjusment} shows that that the optimal convergence rate is again achieved by binary contracts, extending Theorem~\ref{thm:main}, which can be viewed as the special case $T=1$. In the proof (Online Appendix~\ref{app:adjust-pf}), we show that achieving the optimal convergence rate again relies on binary contracts that employ a sequence of maximally lenient tests. As noted in the main text, the convergence rate $\frac{1}{T}\min_{a\in A^-}{\rm KL}(\mu_a, \mu_{a^*})$ is decreasing in the frequency $T$ of the agent's action adjustments, but is the same as in the setting where the agent chooses an action sequence $(a_1, \ldots, a_T)$ without observing the signals generated by past actions.

\subsection{Derivations for Example~\ref{ex:finite-n}}\label{app:illustration}

\noindent{\bf Binary contracts:} For each $n > 2 \ln \frac{2}{1-c}$, consider a binary contract that pays $w^+_n$ if $\overline x \geq \gamma$ and $w^-_n$ otherwise, where $\gamma = 0$ and $w^+_n>w^- _n$ are pinned down by requiring (\ref{eq:IC})--(\ref{eq:IR}) to bind. That is, letting $\Phi$ denote the cdf of the standard Gaussian distribution, we have
\begin{align*}
\underbrace{\left(1 - \Phi[-\sqrt{n}] \right) u(w^+_n) +  \Phi[-\sqrt{n}]  u(w^-_n)}_{\mathbb{E}_1 [u(w(x^n))]} - c = 0 = \underbrace{\frac{1}{2}u(w^+_n) + \frac{1}{2} u(w^-_n)}_{\mathbb{E}_0 [u(w(x^n))]} .
\end{align*}
Thus,
\[
u(w^+_n)=\frac{c}{1-2\Phi[-\sqrt{n}]}, \quad \quad u(w^-_n)= - u(w^+_n)= -\frac{c}{1-2\Phi[-\sqrt{n}]}.
\]
Note that this contract is feasible, i.e., $u(w^-_n), u(w^+_n)\in u([\underline w, \infty))$: Indeed, for all $n$, the Gaussian concentration inequality $\Phi[-\sqrt{n}]\leq \exp[-\frac{n}{2}]$ implies
\begin{equation}\label{eq:gauss-conc}
u(w^+_n)\leq \frac{c}{1-2\exp[-\frac{n}{2}]}.
\end{equation}
Since $n > 2 \ln \frac{2}{1-c}$, (\ref{eq:gauss-conc}) implies $u(w^+_n )<1$, and hence $u(w^-_n) > -1 \geq u(\underline w)$ (by the assumption that $\underline w \leq -\frac{1}{\eta} \ln 2$).

Let $h(v):=u^{-1}(v)=-\frac{1}{\eta}\ln (1-v)$ for all $v<-1$.  Then the difference between the implementation cost under this contract and the first-best cost is 
\begin{eqnarray*}
\left(1-\Phi[-\sqrt{n}]\right) w^++\Phi[-\sqrt{n}] w^- -h(c) 
&\leq& \left(1-\Phi[-\sqrt{n}]\right) \left(h(u (w^+))-h(c)\right) \\
\leq h'(u(w^+)) \left(u (w^+)-c\right) 
&\leq& h'\left(\frac{c}{1-2\exp[-\frac{n}{2}]} \right)  \frac{2c\exp[-\frac{1}{2}n]}{1-2\exp[-\frac{1}{2}n]}\\
&=& \frac{2c}{\eta}\frac{\exp[-\frac{n}{2}]}{1-c-2\exp[-\frac{n}{2}]},
\end{eqnarray*}
where the second and third inequalities use the convexity of $h$ and (\ref{eq:gauss-conc}), and the final equality uses
\[
 h'\left(\frac{c}{1-2\exp[-\frac{n}{2}]} \right)= \frac{1}{\eta}\frac{1-2\exp[-\frac{n}{2}]}{1-c-2\exp[-\frac{n}{2}]}.
\]
This establishes the upper bound in (\ref{eq:bd-bin}) on the difference between the implementation cost under binary contracts and the first-best.

{\bf Linear contracts:} For each $n$, consider a linear contract of the form $w(x^n) = \alpha_n \overline x + \beta_n$. Then, by the properties of log-normal distributions, the agent's expected payoff to choosing action $a$ is
\[
{\mathbb E}_a \left[1-\exp\left[-\eta ( \alpha_n \overline x + \beta_n)\right] \right]-ca=1-\exp[-\eta (\alpha_n a + \beta_n-\frac{\eta \alpha_n^2}{2n})]-ca.
\]
By standard arguments, (\ref{eq:IC})--(\ref{eq:IR}) bind under the optimal $\alpha_n$, $\beta_n$, so the agent's expected payoff from choosing both $a=1$ and $a=0$ must be $0$. For $a = 0$, this yields
\[
1-\exp[-\eta (\beta_n -\frac{\eta \alpha_n^2}{2n})]=0 \iff \beta_n -\frac{\eta \alpha_n^2}{2n}=0.
\]
For $a = 1$, this yields
\[
1-\exp[-\eta (\alpha_n + \beta_n-\frac{\eta \alpha_n^2}{2n})]-c=0 \iff -\frac{1}{\eta} \ln (1-c) =\alpha_n + \beta_n-\frac{\eta \alpha_n^2}{2n}.
\]
Combining yields
\[
\alpha_n =-\frac{1}{\eta}\ln(1-c), \quad \beta_n =\frac{1}{2\eta n} \left(\ln (1-c)\right)^2.
\]
The principal's implementation cost is then
\[
{\mathbb E}_1[w]=\alpha_n + \beta_n=-\frac{1}{\eta}\ln(1-c) + \frac{1}{2\eta n} \left(\ln (1-c)\right)^2.
\]
Thus, the difference with $C^{\rm FB} =  -\frac{1}{\eta}\ln (1-c)$ is $ \frac{1}{2\eta n} \left(\ln (1-c)\right)^2$, as claimed in (\ref{eq:bd-lin}).

\footnotesize
\begin{spacing}{0.01}
\bibliographystyle{econometrica}
\bibliography{moral-hazard-0701}
\end{spacing}

%%%%%%%%%%%
\newpage
\setcounter{page}{1}

\singlespacing

\begin{center}
 {\Large\textbf{Online Appendix to ``Monitoring with Rich Data''\\[0.7cm]}}
 {\large Mira Frick, Ryota Iijima, and Yuhta Ishii\\[1cm]}
\end{center}

\normalsize

\section{Proofs for Appendix~\ref{app:general}}

\subsection{Proof of Theorem~\ref{thm:general}}\label{app:noniid-pf}

\subsubsection{Properties of Rate Functions}

Let $I_{a, a'}$ denote the rate functions associated with sequence $(\mu^n)$.

\begin{lem}\label{lem:consistency}For each $a\not=a^*$ and $\ell\in\mathbb R$, we have:
\begin{enumerate} 
\item $I_{a^*,a}(\ell)-I_{a, a}(\ell)=- \ell$.

\item $I_{a^*,a}( \ell)\geq I_{a^*,a}(\hat L_a(a))$ if  $\ell\leq \hat L_a(a)$.
 
\end{enumerate}

\end{lem}

\begin{proof} {\bf \underline{First part}:} For any $\varepsilon>0$, let $\Gamma_\varepsilon:=[\ell-\varepsilon, \ell+\varepsilon]$. By the mediant inequality,
\[
\ell-\varepsilon\leq \frac{1}{n}\ln \frac{{\mathbb P}_{a^*}[ L_n(a)\in\Gamma_\varepsilon]}{{\mathbb P}_a [ L_n(a)\in \Gamma_\varepsilon ]}\leq  \ell+\varepsilon.
\]
Fix any $\varepsilon'>0$. Since $I_{a^*,a}$ is lower-semicontinuous,  by choosing $\varepsilon$ sufficiently small, we have
$I_{a^*, a}(\ell')\geq I_{a^*, a}(\ell)-\varepsilon'$ for all $\ell'\in \Gamma_\varepsilon$.  Then
\begin{eqnarray*}
\lim_{n\to\infty}\frac{1}{n}\ln \frac{{\mathbb P}_{a^*}( L_n(a)\in  \Gamma_\varepsilon)}{{\mathbb P}_{a}( L_n(a)\in  \Gamma_\varepsilon)} 
&=&-\inf_{ \ell'\in \Gamma_\varepsilon}I_{a^*,a}( \ell')+\inf_{ \ell'\in  \Gamma_\varepsilon}I_{a, a}( \ell') \\
&\leq& - I_{a^*,a}( \ell)+\varepsilon'+I_{a, a}( \ell),
\end{eqnarray*}
where the equality uses (\ref{eq:LDP}). Thus,  $ \ell-\varepsilon\leq  -I_{a^*,a}( \ell)+\varepsilon'+I_{a,a}( \ell)$. Since $\varepsilon,\varepsilon'$ can be chosen arbitrarily small, we have $ \ell\leq  - I_{a^*,a}( \ell)+I_{a,a}( \ell)$.  A symmetric argument yields $\ell\geq  - I_{a^*,a}( \ell)+I_{a,a}( \ell)$. 

{\bf \underline{Second part}:} Observe that for any $\ell\leq \hat L_a(a)$ with $I_{a^*,a}(\ell)<\infty$, we have
\[
I_{a^*,a}( \ell)=I_{a,a}( \ell)- \ell \geq I_{a,a}(\hat L_a(a))-\hat L_a(a)=I_{a^*,a}(\hat L_a(a)),
\]
where the two equalities use the first part and the inequality uses $I_{a,a}(\hat L_a(a))=0\leq I_{a,a}(\ell)$. 
\end{proof}

\subsubsection{Upper Bound on Inefficiency under Binary Contracts}

%%%%%%%%%%%%%%%%%%%%%%%%%

We first show that, as $n\to\infty$,  $C^{\rm bin}_n(\mu, u, c, a^*)-C^{\text{FB}}(u, c, a^*)$ vanishes at least as fast as at the exponential rate $\rho(\mu)$.

By the first part of Lemma~\ref{lem:consistency} and the identification condition, we have $-\hat L_a(a)=I_{a^*, a}(\hat L_a(a))-I_{a,a}(\hat L_a(a))=I_{a^*,a}(\hat L_a(a))>0$ and $-\hat L_{a^*}(a)=I_{a^*,a}(\hat L_{a^*}(a))-I_{a,a}(\hat L_{a^*}(a))=-I_{a,a}(\hat L_{a^*}(a))<0$ for each $a\not=a^*$. By the choice of $\gamma(\cdot)$, for each $a\in A^-$,
\[
\inf_{ \ell< \gamma(a)} I_{a^*,a}(\ell) >0 \text{ and } \inf_{\ell\geq \gamma(a)} I_{a,a}(\ell)>0.
\]
Thus, by (\ref{eq:LDP}), 
\begin{equation}\label{eq:no error}
\lim_{n\to\infty}p_{a^*, n}=1  \text{ and } \lim_{n\to\infty}p_{a, n}=0  \text{ for each } a\in A^-. 
\end{equation}
As in the proof of Theorem~\ref{thm:main}, we can use (\ref{eq:no error}) to construct binary test contracts $(v_n)$ with thresholds $\gamma(a)$ that satisfy (\ref{eq:IR''}) with equality and (\ref{eq:IC''}) for large enough $n$. Note that
\begin{eqnarray*}
\lim_{n\to\infty}\frac{1}{n}\ln(1-p_{a^*, n})&=& \lim_{n\to\infty}\frac{1}{n}\ln({\mathbb P}_{a^*}[ L_n(a)<\gamma(a) \text{ for some } a\in A^-]) \\
&=& \max_{a\in A^-}\lim_{n\to\infty}\frac{1}{n}\ln({\mathbb P}_{a^*}[ L_n(a)<\gamma(a) ]) \\
&\leq& -\min_{a\in A^-} \inf_{ \ell\leq\gamma(a)}I_{a^*,a}( \ell). %= -\min_{a\in A^-} I_{a^*, a}(\gamma(a)),
\end{eqnarray*}
Thus, as in the proof of Theorem~\ref{thm:main},
\begin{eqnarray*}
C^{\rm bin}_n(\mu, u, a^*, c)-C^{\text{FB}}(u,a^*, c) &\leq& \mathbb E_{a^*}[h(v_n(x^n))]-h(c(a^*)) \\
 &\leq& \exp[- n\min_{a\in A^-} \inf_{\ell\leq\gamma(a)}I_{a^*,a}(\ell)  +o(n)].
\end{eqnarray*}
Finally, note that for each $a\in A^-$,
\[
\sup_{\gamma(a)>\hat L_a(a)}\inf_{ \ell\leq\gamma(a)}I_{a^*,a}(\ell)=\inf_{\ell \leq \hat L_a(a)}I_{a^*,a}(\ell)=I_{a^*,a}(\hat L_{a}(a)),
\]
where the second equality uses the second part of Lemma~\ref{lem:consistency} and the local continuity of $I_{a^*, a}$. Thus, by choosing $\gamma(a)$ arbitrarily close to $\hat L_{a}(a)$ for each $a\in A^-$, it follows that  $C^{\rm bin}_n(\mu, u, c, a^*)-C^{\text{FB}}(u, c, a^*)$ vanishes at least as fast as at the exponential rate $\min_{a\in A^-}I_{a^*, a}(\hat L_{a}(a))=\rho(\mu)$.

\subsubsection{Lower Bound on Inefficiency under General Contracts}
Next, we show that, as $n\to\infty$,   $C^{\rm{SB}}_n(\mu, u,c, a^*)-C^{\text{FB}}(u, c, a^*)$ vanishes no faster than at the exponential rate $\rho(\mu)$.

Let $\hat a \in \argmin_{a\in A^-}I_{a^*,a}(\hat L_{a}(a))$, i.e., $\rho(\mu)=I_{a^*,\hat a}(\hat L_{\hat a}(\hat a))$.  We consider a relaxed problem with action set $\{a^*, \hat a\}$.  
As in the proof of Theorem~\ref{thm:main}, it is sufficient to show that the difference between the second-best cost  under the relaxed problem and $C^{\text{FB}}(u,c, a^*)$ cannot vanish  faster than at the exponential rate $I_{a^*, \hat a}(\hat L_{\hat a}(\hat a))$.

As in the proof of Theorem~\ref{thm:main}, we consider a sequence of optimal contracts $(v_n)$ for the relaxed problem, where each $v_n$ is written as a weakly increasing function of $L_n$. The sequence is uniformly bounded by the same argument as in Appendix~\ref{app:optimal}. 

Lemma~\ref{lem:flat} from the proof of Theorem~\ref{thm:main} and its proof generalizes to the current setting by using the general large-deviation principle (\ref{eq:LDP}) instead of Cram\'er's theorem. Lemma~\ref{lem:IC bound2} below generalizes Lemma~\ref{lem:IC bound} to the current setting. Given this, the remaining proof follows the same arguments as the proof of Theorem~\ref{thm:main}.

 \begin{lem}\label{lem:IC bound2}
 Consider any uniformly bounded sequence $(v_n)$ of contracts satisfying (\ref{eq:IC''}) under the relaxed problem. Then ${\rm Var}_{a^*}[v_n( L_n)] \geq \exp[-I_{a^*, a}(\hat L_{\hat a}(\hat a)) n+o(n)]$. \end{lem}

\begin{proof}
Suppose to the contrary that ${\rm Var}_{a^*}[v_n( L_n)] \leq \exp[-\lambda n+o(n)]$ for some $\lambda> I_{a^*, \hat a}(\hat L_{\hat a}(\hat a))$. By local continuity of $I_{a^*, \hat a}$, there is $\underline\gamma<\hat L_{\hat a}(\hat a)$ with $I_{a^*, \hat a}(\underline\gamma)<\lambda$. Moreover, by local continuity of $I_{a^*, \hat a}$ and the fact that $I_{a^*\hat a}(\hat L_{a^*}(\hat a))=0$, there is $\overline\gamma>\hat L_{a^*}(\hat a)$ with $I_{a^*,\hat a}(\overline\gamma)<\lambda$. By (the generalized) Lemma~\ref{lem:flat}, $\lim_{n\to \infty} \left(v_n(\overline\gamma)-v_n(\underline\gamma) \right)=0$.

At the same time, for each $a \in \{a^*,\hat a\}$, the fact that $I_{a,\hat a}$ achieves 0 only at $\hat L_a(\hat a)\in (\underline\gamma, \overline\gamma)$ and that the lower level sets are compact implies $\inf_{\gamma\in (-\infty, \underline\gamma]\cup [\overline\gamma, \infty)}I_{a,\hat a}(\gamma)>0$. Thus, ${\mathbb P}_a[L_n(\hat a)\in [\underline\gamma,\overline\gamma]]\to 1$ by (\ref{eq:LDP}).  As the sequence of contracts is uniformly bounded, this implies $\lim_{n\to\infty} \left( {\mathbb E}_{a^*}[v_n(L_n)]-{\mathbb E}_{\hat a}[v_n(L_n)] \right)=0$. Then (\ref{eq:IC''}) is violated for all large enough $n$, a contradiction. 
\end{proof}

\subsection{Proof of Theorem~\ref{prop:adjusment}}\label{app:adjust-pf}

\subsubsection{Upper Bound on Inefficiency under Binary Contracts}

We first show that, as $n\to\infty$,  $G^{\rm FB} -G^{\text{bin}}_n$ vanishes at least as fast as at the exponential rate $\frac{1}{T}\min_{a\in A^-}{\rm KL}(\mu_a, \mu_{a^*})$.

For each $a\not=a^*$ and $t=1,\ldots, T$, define
\[
 L_{t, n}(a):= \frac{T}{n}\sum_{i=1}^{\frac{n}{T}}\ln\frac{d\mu_{a^*}}{d\mu_{a}}(x_{t,i}).  
\]
For each $a \in A^-$, fix some $\gamma(a) \in ({\mathbb E}_a[ L_{t, n}(a)],{\mathbb E}_{a^*}[ L_{t,n}(a)])$; note that the expectations ${\mathbb E}_{a'} [ L_{t, n}(a)]$ do not depend on $n$ and $t$. For each $a' \in A$, define
\[
p_{a', n}:={\mathbb P}_{a'}[ L_{t,n}(a)\geq \gamma(a), \, \forall a\in A^-, t=1,\ldots, T].
\]
By the weak law of large numbers and the choice of $\gamma(a)$, we have
\begin{equation}\label{eq:LLN}
\lim_{n\to\infty}p_{a^*, n}=1  \text{ and } \lim_{n\to\infty}p_{a, n}=0  \text{ for all } a\in A^- .
\end{equation}

Consider the recommended strategy $\alpha^*$ given by $\alpha^*_1 \equiv a$ and, for all $t = 2, \ldots, T$,
\[
\alpha^*_t (x^n_1,\ldots, x^n_{t-1})= \begin{cases}
a^* &\text{ if }   L_{\tau,n}(a)\geq \gamma(a)   \text{ for all  } a\in A^-, \tau=1,\ldots, t-1 \\
\underline a &\text{ otherwise } 
\end{cases}
\]
for some $\underline a\in\arg\min_{a\in A}c(A)$. Consider the sequence of binary contracts $(v_n)$ of the form 
\[
v_n(x^n)= \begin{cases}
v^+_n & \text{ if }   L_{k,n}(a)\geq \gamma(a)   \text{ for all  } a\in A^-, k=1, \ldots, T \\
v^-_n & \text{ otherwise.} 
\end{cases}
\]
Here
\[
v^+_n:= \frac{(1-\overline p_{n}p^{T-1}_{a^*,n})\beta^*_n-(1-p_{a^*, n}^{T})\beta_n}{p^{T-1}_{a^*,n}(p_{a^*, n}-\overline p_{n})}, \quad
 v_n^-:=\frac{p_{a^*, n}^{T}\beta_n-\overline p_{n}p^{T-1}_{a^*,n}\beta^*_n }{p^{T-1}_{a^*,n}(p_{a^*, n}-\overline p_{n})},
\]
where
\[
\overline p_n:=\max_{a\in A^-}p_{a, n}, \quad \beta^*_n:={\mathbb E}_{\alpha^*}\left[\frac{\sum_{t=1}^Tc(a_t)}{T}\right], \quad \beta_n:=\min_{a\in A^-}{\mathbb E}_{\alpha^*_a}\left[\frac{\sum_{t=1}^Tc(a_t)}{T}\right],
\]
and $\alpha^*_a$ denotes the strategy defined by a one-shot deviation from $\alpha^*$ to $a\not=a^*$ at $t=1$. Note that, by (\ref{eq:LLN}), 
\begin{equation}\label{eq:v-T}
\lim_{n\to\infty} v^+_n=\lim_{n\to\infty}\beta^*_n=c(a^*) \text{ and } \lim_{n\to\infty} v^-_n=\lim_{n\to\infty}\beta_n=c(\underline a),
\end{equation}
so $v^+_n$, $v^-_n \in u([\underline w, \infty))$ for all sufficiently large $n$. Moreover, contracts $(v_n)$ satisfy
\[
p^T_{a^*, n}v_n^++(1-p^T_{a^*, n})v_n^- -\beta^*_n =0,
\]
\[ p^T_{a^*, n}v_n^++(1-p^T_{a^*, n})v_n^- -\beta^*_n
=\overline p_{n}p^{T-1}_{a^*, n}v_n^++(1-\overline p_{n}p^{T-1}_{a^*, n})v_n^- -\beta_n.
 \]
 The first line implies that (\ref{eq:IR-adj}) binds under strategy $\alpha^*$, as the LHS is the agent's expected payoff under $\alpha^*$. The second line ensures that one-shot deviations from $\alpha^*$ to any action in $A^-$ at $t=1$ are not strictly profitable for the agent.

Conditional on any history at which $\alpha^*$ prescribes $\underline a$, the wage becomes constant, so the agent has no incentive to deviate. Let $U_{n, t}(a)$ denote the agent's expected total payoff from a one-shot deviation from $\alpha^*$ to $a\not=a^*$ at some $t$ and history $(x^n_1,\ldots, x^n_{t-1})$ at which $\alpha^*$ prescribes $a^*$. (Note that this value is independent of the particular history). Observe that, absent any deviation, the agent's expected total payoff conditional on such histories tends to 0 as $n\to\infty$ by (\ref{eq:LLN})-(\ref{eq:v-T}) and binding (\ref{eq:IR-adj}), where the convergence is uniform across histories.

For any $a\in A^-$ and $t\geq 2$, (\ref{eq:LLN})-(\ref{eq:v-T}) imply
\[
\lim_{n\to\infty}U_{n,t}(a)=\frac{tc(\underline a)-c(a)-(t-1)c(a^*)}{T}<0,
\]
where the inequality follows from $c(\underline a)\leq c(a)<c(a^*)$. Thus, there is no incentive to deviate to actions in $A^-$ at large enough $n$. 
For any $a\in A^+$ and $t\geq 1$, (\ref{eq:LLN})-(\ref{eq:v-T}) imply
\begin{eqnarray*}
\limsup_{n\to\infty}U_{n,t}(a)&=&\limsup_{n\to\infty} p_{a,n} \frac{c(a^*)-c(a)}{T}+(1-p_{a,n})\frac{tc(\underline a)-c(a)-(t-1)c(a^*)}{T}<0,
\end{eqnarray*}
where the inequality follows from $c(\underline a)<c(a^*)<c(a)$. Thus, there is no incentive to deviate to actions in $A^+$ at large enough $n$.

As in the proof of Theorem~\ref{thm:main}, we can show that $1-p_{a^*, n}=\exp[- \frac{n}{T}\min_{a\in A^-} \inf_{ \ell<\gamma(a)}I_{a^*, a}(\ell) +o\left(\frac{n}{T}\right)]$. Moreover, observe that $1-p^T_{a^*, n}$ and $\beta^*_n-c(a^*)$ also vanish at the same rate as $1-p_{a^*, n}$. Thus, 
\begin{eqnarray*}
&&G^{\rm FB}-G^{\rm bin}_n  \\
&\leq& \sum_{t=1}^{T-1}p_{a^*, n}^{t-1}(1-p_{a^*, n})\frac{t (g(a^*)-g(\underline a))}{T} + p^T_{a^*, n} \left(h\left(v_n^+\right)-h (c(a^*))\right)\\
&&+(1-p^T_{a^*, n})  \left(h\left(v_n^-\right)- h (c(a^*)) \right) \\
 &\leq&\sum_{t=1}^{T-1}p_{a^*, n}^{t-1}(1-p_{a^*, n})\frac{t (g(a^*)-g(\underline a))}{T} + p^T_{a^*, n} \left(h\left(v_n^+\right)-h (c(a^*))\right) \\
 &=& \exp[- \frac{n}{T} \min_{a\in A^-} \inf_{ \ell<\gamma(a)}I_{a^*,a}(\ell) +o\left(\frac{n}{T}\right)],
\end{eqnarray*}
where the last line follows from observing that 
\[
v_n^+-c(a^*)=\frac{(1-p^T_{a^*,n})(\beta_n^*-\beta_n)}{p^{T-1}_{a^*,n}(p_{a^*,n}-\overline p_n)}+\beta^*_n-c(a^*)
\]
vanishes at the same exponential rate as  $1-p_{a^*, n}$.

The remaining step of letting $\gamma(a)\searrow{\mathbb E}_{a^*}[ L_{t, n}(a)]$ for each $a\in A^-$ is analogous to the proof of Theorem~\ref{thm:main}.

\subsubsection{Lower Bound on Inefficiency under General Contracts}

Next, we show that, as $n\to\infty$, $G^{\rm FB} -G^{\text{SB}}_n$ vanishes no faster than at the exponential rate $\frac{1}{T}\min_{a\in A^-}{\rm KL} (\mu_a, \mu_{a^*})$.

By standard arguments, $G^{\rm SB}_n$ can be written in a recursive manner by taking the agent's continuation value as a state variable:  Let $G^{\rm SB}_{n, T+1}(z):=-h(z)$ for each $z\in u([\underline w,\infty))$. For each $t=T, T-1,\ldots, 1$ and $z\in u([\underline w, \infty))-\frac{T-t+1}{T}c(A)$, we inductively define  $G^{\rm SB}_{n, t}(z):=\max_{a\in A}G^{\rm SB}_{n, t}(z, a)$, where for each $a\in A$,
\[
G^{\rm SB}_{n, t}(z, a):=\sup_{\psi: X^{\frac{n}{T}}\to u([\underline w,\infty))-\frac{T-t}{T}c(A)} \frac{b(a)}{T} +{\mathbb E}_{a}[ G_{n, t+1}(\psi(x^n_t))]
\]
subject to IC 
\begin{equation}\label{eq:IC1}
-\frac{c(a)}{T}+{\mathbb E}_{a}[\psi(x^n_t)] \geq -\frac{c(a')}{T}+{\mathbb E}_{a'}[\psi(x^n_t)]  \; \forall a'\in A
\end{equation}
and promise keeping 
\begin{equation}\label{eq:IR1}
-\frac{c(a)}{T}+{\mathbb E}_{a}[\psi(x^n_t)] = z,
\end{equation}
where $\psi(\cdot)$ specifies the agent's continuation value as a function of the realized $x^n_t$.\footnote{The existence of an optimal policy in this problem is guaranteed by analogous arguments as in Appendix~\ref{app:optimal} when $n$ is sufficiently large.} 
The optimal value of the principal is then given by $G^{\rm SB}_n=G^{\rm SB}_{n,1}(0)$.

The first-best value can be written analogously:  Let $G^{\rm FB}_{n, T+1}(z):=-h(z)$ for each $z\in u([\underline w,\infty))$. For each $t=T, T-1,\ldots, 1$ and $z\in u([\underline w, \infty))-\frac{T-t+1}{T}c(A)$, we inductively define $G^{\rm FB}_{n, t}(z):=\max_{a\in A}G^{\rm FB}_{n, t}(z, a)$, where for each $a\in A$,
\[
G^{\rm FB}_{n, t}(z, a):=\sup_{\psi: X^{\frac{n}{T}}\to u([\underline w,\infty))-\frac{T-t}{T}c(A)} \frac{b(a)}{T} +{\mathbb E}_{a}[ G^{\rm FB}_{n, t+1}(\psi(x^n_t))]
\]
subject to promise keeping (\ref{eq:IR1}). Then $G^{\rm FB}=G_{n, 1}^{\rm FB}(0)$. 
By the richness condition on $\mu_a$, in the above problem the principal can induce any randomization over continuation values subject to promise keeping by appropriately choosing $\psi$. Thus, by standard arguments for information design problems,
\[
G^{\rm FB}_{n, t}(z, a)=\frac{b(a)}{T}+ {\rm cav}G^{\rm FB}_{n, t+1}\left(z+\frac{c(a)}{T}\right),
\]
where ${\rm cav}$ denotes the concavification operator.  Consequently, $G^{\rm FB}_{n, t}$ is independent of $n$, so we drop the subscript $n$ and simply write $G^{\rm FB}_{t}$.

As we show in Appendix~\ref{app:FB-bound} below, there is a $C^2$ function $f: u([\underline w,\infty))-\frac{(T-1)c(A)}{T}\to \mathbb R$ that upper-bounds $G^{\rm FB}_2$ and such that $f'<0$, $f''<0$,  $\inf_wf'(w)=-\infty$, and $f(\frac{c(a^*)}{T})= \frac{(T-1)b(a^*)}{T}+h(c(a^*))$.

Since $f(\cdot)\geq G^{\rm SB}_{n,2}(\cdot)$, we have that $G^{\rm SB}_{n, 1}(0, a^*)$ is weakly lower than the value of the problem 
\begin{equation}\label{eq:relax2}
\max_{\psi: X^{\frac{n}{T}}\to u([\underline w, \infty))-\frac{T-1}{T}c(A)} \frac{b(a^*)}{T} +{\mathbb E}_{a^*}[f(\psi(x^n_1))] = \frac{b(a^*)}{T}-\min_{\psi: X^{\frac{n}{T}}\to u([\underline w, \infty))-\frac{T-1}{T}c(A)}{\mathbb E}_{a^*}[-f(\psi(x^n_1))] 
\end{equation}
subject to (\ref{eq:IC1}) and (\ref{eq:IR1}) with $a=a^*$ and $z=0$.

Observe that the minimization problem in (\ref{eq:relax2}) can be viewed as the second-best problem of our baseline model in Section~\ref{sec:model},  by treating $\frac{c(\cdot)}{T}$ and $-f(\cdot)$ as the agent's cost function and inverse utility function, respectively. Thus, by Theorem~\ref{thm:main}, the difference between $\frac{b(a^*)}{T}+f(\frac{c(a^*)}{T})=b(a^*)+h(c(a^*))$ and the value in (\ref{eq:relax2}) is $\exp[ -\frac{n}{T}\min_{a\in A^-}{\rm KL}(\mu_a, \mu_{a^*}) + o(\frac{n}{T})]$. Therefore, $G^{\rm SB}_{n, 1}(0, a^*)$ cannot converge to $b(a^*)+h(c(a^*))=G^{\rm FB}$ faster than at rate $\frac{1}{T}\min_{a\in A^-}{\rm KL}(\mu_a, \mu_{a^*})$.

Finally, for any $a\not=a^*$, we have $G^{\rm FB}_{1}(0, a)<G^{\rm FB}_{1}(0, a^*)=G^{\rm FB}$
by the assumption that it is strictly optimal to implement $a_1=a^*$ under the first-best problem. Thus, $G^{\rm SB}_{n, 1}(0, a)$, which is weakly lower than $G^{\rm FB}_{1}(0, a)$, is bounded away from the first-best $G^{\rm FB}$. Combined with the above observations, this shows that $G^{\rm SB}_{n}=\max_{a\in A}G^{\rm SB}_{n, 1}(a^*)$ cannot converge to $b(a^*)+h(c(a^*))=G^{\rm FB}$ faster than at rate $\frac{1}{T}\min_{a\in A^-}{\rm KL}(\mu_a, \mu_{a^*})$.

\subsubsection{Details of Bounding $G^{\rm FB}_2$}\label{app:FB-bound}

Since by construction ${\rm cav}G^{\rm FB}_2$ is a concave and decreasing function,it is dominated by some linear and decreasing function. To ensure that the bound $f$ can be chosen to be appropriately concave, we make some preliminary observations.

Because it is strictly optimal under the first-best problem to implement $a^*$ at all $t$ and $\frac{(t-1)c(a^*)}{T}$ is the agent's (on-path) continuation value,  we have 
\[
{\rm cav}G^{\rm FB}_{t}\left(\frac{(t-1)c(a^*)}{T}+\varepsilon\right)=G^{\rm FB}_{t}\left(\frac{(t-1)c(a^*)}{T}+\varepsilon\right)=\frac{(T-t+1)g(a^*)}{T}-h(c(a^*)+\varepsilon)
\] 
whenever $|\varepsilon|$ is sufficiently small. This in particular implies that ${\rm cav}G^{\rm FB}_{2}\left(\frac{c(a^*)}{T}\right)=\frac{(T-1)g(a^*)}{T}-h(c(a^*)$, and hence 
\begin{equation}\label{eq:cavFB}
{\rm cav}(G^{\rm FB}_{2})'\left(\frac{c(a^*)}{T}\right)=-h'(c(a^*)<0 \quad \text{ and } \quad {\rm cav}(G^{\rm FB}_{2})''\left(\frac{c(a^*)}{T}\right)=-h''(c(a^*)<0. 
\end{equation}

Let $\hat a$ be an action that maximizes $b(\cdot)$ among $\arg\min_{a\in A}c(a)$. 
Since $h'(z)\to\infty$ as $z\to \sup_{w}u(w)$,  we have ${\rm cav}G^{\rm FB}_T(z)=G^{\rm FB}_T(z)= \frac{g(\hat a)}{T}-h(z+\frac{c(\hat a)}{T})$ for all large enough $z$. By induction, for each $t=T, T-1,\ldots, 1$ and sufficiently large $z$,
\[
{\rm cav}G^{\rm FB}_t(z)=G^{\rm FB}_t(z)= \frac{(T-t+1)g(\hat a)}{T}-h(z).
\] 
Thus,
\begin{equation}\label{eq:cavFB2}
{\rm cav}(G^{\rm FB}_2)'(z)\to -\infty 
\end{equation}
as $z$ tends to the upper-bound of the range of ${\rm cav}(G^{\rm FB}_2)$. 
Then based on (\ref{eq:cavFB})-(\ref{eq:cavFB2}), there is a $C^2$ function $f$ that upper-bounds $G^{\rm FB}_2$ and such that $f'<0$, $f''<0$,  $\inf_wf'(w)=-\infty$, and $f(\frac{c(a^*)}{T})= G^{\rm FB}_2(\frac{c(a^*)}{T})=\frac{(T-1)b(a^*)}{T}+h(c(a^*))$.

 \end{document}